\documentclass[gmd, manuscript]{copernicus}

\usepackage{graphicx}
\usepackage{subfig}

\usepackage{siunitx}
\DeclareSIUnit[number-unit-product = \,]{\Phosphat}{P}
\DeclareSIUnit[number-unit-product = \,]{\Modelyear}{yr}
\DeclareSIUnit[number-unit-product = \,]{\Timestep}{dt}

\begin{document}
\nolinenumbers

\title{Shortening the runtime using larger time steps for the simulation of 
marine ecosystem models}

\Author[1]{Markus}{Pfeil}
\Author[1]{Thomas}{Slawig}

\affil[1]{Kiel Marine Science -- Centre for Interdisciplinary Marine Science,
          Dep. of Computer Science, Kiel University, 24098 Kiel, Germany}

\runningtitle{Shortening the simulation runtime of marine ecosystem models}
\runningauthor{Pfeil and Slawig}
\correspondence{Markus Pfeil (mpf@informatik.uni-kiel.de)}

\received{}
\pubdiscuss{} 
\revised{}
\accepted{}
\published{}


\firstpage{1}

\maketitle

\begin{abstract}
  The reduction of computational costs for marine ecosystem models is important
  for the investigation and detection of the relevant biogeochemical processes
  because such models are computationally expensive. In order to lower these
  computational costs by means of larger time steps we investigated the accuracy
  of steady annual cycles (i.e., an annual periodic solution) calculated with
  different time steps. We compared the accuracy for a hierarchy of
  biogeochemical models showing an increasing complexity and computed the steady
  annual cycles with offline simulations that are based on the transport matrix
  approach. For each of these biogeochemical models, we obtained practically the
  same solution even though larger time steps. This indicates that larger time
  steps shortened the runtime with an acceptable loss of accuracy.
\end{abstract}


\introduction 
  Shortening the runtime for simulations of marine ecosystem models is important
for computing steady annual cycles. In general, marine ecosystem models are an
essential element to investigate the influence of various biogeochemical
processes in the marine carbon cycle. The ocean biota, for example, processes
different climatically relevant chemical elements and the ocean takes up carbon
from the atmosphere and stores it. Due to the interplay of physical and
biogeochemical processes, a marine ecosystem model consists of a global ocean
circulation model coupled to a biogeochemical model \citep[cf.][]{Fasham03,
SarGru06}. While the equations and variables of the physical processes are well
known, a set of state variables or equations describing the biogeochemical
processes is generally not available. Therefore, there is a wide range of
biogeochemical models which differ in their complexity due to the number of
state variables and parameterizations \citep[see e.g.,][]{KrKhOs10, KeOsEb12,
ISSMLN13, YoPoAn13, LBMAAB16}. The validation of these models requires many
calculations of steady annual cycles. More specifically, the validation, which
contains a parameter estimation and a discussion of simulation results, assesses
the steady annual cycles against given observational data \citep{FLSW01}.

The computational effort is tremendous for any fully coupled simulation because
a single model evaluation is already computationally expensive \citep{Osc06}.
This results from the simultaneous computation of the ocean circulation and the
biogeochemical model in three spatial dimensions. In particular, the
computational cost increases for the computation of a steady annual cycle due to
the necessary long-time integration \citep[cf.][]{BeDiWu08}. In typical cases,
the simulation needs at least several thousand model years to reach a steady
annual cycle \citep[cf.][]{Bryan84, DaMcLa96, BeDiWu08}.

Several strategies address the reduction of the computational costs in order
to compute a steady annual cycle in marine ecosystem models
\cite[e.g.][]{Bryan84, DaMcLa96, Wang01, SiPiSl13}. Firstly, spatial
parallelization using domain decomposition methods lowers the computational
costs. Instead of using the fully coupled simulation (the so-called
\emph{online} simulation), the \emph{offline} simulation, secondly, neglects
the influence of the biogeochemical model on the ocean circulation and, hence,
applies a pre-computed ocean circulation. A third strategy is the use of
Newton's method instead of the long-time integration. Lastly, the application of
graphics processing units shortens the computational time. This constitutes only
an excerpt of the strategies for the reduction of the computational costs.

\citet{KhViCa05} reduced the computational effort with a tolerable loss of
accuracy by implementing the transport matrix method (TMM) as offline simulation,
which approximated the general ocean circulation by pre-computed transport
matrices. On the one hand, this reduced the computation of the ocean
circulation to matrix-vector multiplications and, on the other hand, separated
the evaluation of the biogeochemical model from the ocean circulation
neglecting the very small effects of biogeochemical tracers on the density
(e.g., via solar heating) \citep[][]{Osc04}. \citet{Kha07} showed that the
accuracy of this method using the linearized tracer transport is
sufficient for marine ecosystem models to obtain first insights on a global
basin scale. The method also provides the flexibility to replace the long-time
integration with Newton's method \citep{Kha08}.

The computational time to compute a steady annual cycle strongly affects the
computational effort of a parameter optimization or sensitivity study, since
several hundred computations of steady annual cycles are necessary. With the
help of parameter optimization of global biogeochemical models, the optimal
model parameters are determined so that the model ideally reflects the
real-world data \citep{KwoPri06, KwoPri08, PPKOS13}. This is especially
necessary to optimize the poorly known model parameters.

In order to lower the computational costs, we have investigated, in this paper,
the influence of larger time steps on the accuracy of the computation of steady
annual cycles for both the biogeochemical model \citep{DuSoScSt05} which is part
of the MITgcm ocean model and a hierarchy of biogeochemical models with an
increasing complexity \citep{KrKhOs10}. To our knowledge, there is no
systematic investigation of this influence in the literature. However, this
is essential for the selection of an appropriate time step to lower the
computational effort of those computations. Indeed, some studies had already
used larger time steps \citep[e.g.,][]{PPKOS13, KSKSO17} and, in particular, the
use of larger time steps shortened the runtime with an acceptable loss of
accuracy.

This paper is organized as follows: after an introduction to the general
structure of marine ecosystem models, we describe in Sect.
\ref{sec:ModelDescription} the TMM and the computation of a steady annual
cycle. In Sect. \ref{sec:Timesteps}, we present the computation using larger
time steps. Section \ref{sec:Results} covers numerical results in which
different time steps were used. The paper closes with conclusions deduced from
the numerical results in Sect. \ref{sec:Conclusion}.

\section{Model description}
\label{sec:ModelDescription}


A marine ecosystem model usually consists of a component describing the ocean
circulation and a component representing the biogeochemical model
\citep[e.g.,][]{Fasham03, FenNeu04, SarGru06}. Biogeochemical tracers are
substances in the ocean water that are subject to chemical or biochemical
reactions. The equations modeling the ocean circulation including
temperature and salinity distribution are coupled to equations regulating the
transport and the reactions of the biogeochemical tracers. The couplings are
based on the fact that the ocean circulation affects the tracer concentrations
and the turbulent mixing of marine water dominates the diffusion of the tracer
concentrations and, vice versa, the tracer concentrations influence the ocean
circulation. However, a fully coupled simulation (also called \emph{online}
model) in three spatial dimensions is limited to single model evaluations, even
on high performance computing clusters, because the simulation of both systems
must be carried out simultaneously and the computational effort, therefore, is
enormous.

Instead of the online model, the \emph{offline} model simplifies the simulation.
The offline simulation uses passive tracers which do not have an effect on the
ocean physics or neglects this impact. Consequently, the coupling is only a
one-way coupling from the ocean circulation to the tracer dynamics. Due to the
one-way coupling for offline simulations, we can apply a pre-computed ocean
circulation. We applied offline models with an increasing complexity of the
biogeochemical models as marine ecosystem models, thus they were introduced by
\citet{KrKhOs10, DuSoScSt05}.

\subsection{Model equations for marine ecosystems}
\label{sec:Model-ModelEquationsMarineEcosystems}

  A system of differential equations describes the marine ecosystem model. The
  differential equations are of the Lotka-Volterra or predator-prey type, and
  the number of tracers defines the size of the system of differential equations
  \citep{Lot10, Vol31}. We consider, in the rest of this paper, marine ecosystem
  models using an offline model with $n_{y} \in \mathbb{N}$ tracers on a spatial
  domain $\Omega \subset \mathbb{R}^3$ (i.e., the ocean) and on a time interval
  $[0,1]$ (i.e., one model year). For $i \in \left\{1, \ldots, n_y \right\}$,
  $y_i: \Omega \times [0,1] \rightarrow \mathbb{R}$ denotes the function of the
  tracer concentration and $\vec{y} := \left( y_i \right)_{i=1}^{n_{y}}$ the
  vector of all tracers. The following system of parabolic partial differential
  equations describes the tracer transport of a marine ecosystem model
  \begin{align}
  \label{eqn:Modelequation}
      \frac{\partial y_i}{\partial t} (x,t)
           + \left( D (x,t) + A(x,t) \right) y_i (x,t)
        &= q_i \left( x, t, \vec{y}, \vec{u} \right),
        & x \in \Omega, t &\in [0,1], \\
    \label{eqn:Boundarycondition}
      \frac{\partial y_i}{\partial n} (x,t) &= 0,
        & x \in \partial \Omega, t &\in [0,1],
  \end{align}
  for $i = 1, \ldots, n_{y}$, including linears operators $D: \Omega \times
  [0,1] \rightarrow \mathbb{R}$ and $A: \Omega \times [0,1] \rightarrow
  \mathbb{R}$, which corresponds to the diffusion and advection coming from the
  ocean circulation, and the term $q_i: \Omega \times [0,1] \rightarrow
  \mathbb{R}, \left( x, t \right) \mapsto q_i \left( x, t, \vec{y}, \vec{u}
  \right)$ for the tracer $y_i$ of the biogeochemical model. The Neumann
  boundary conditions \eqref{eqn:Boundarycondition} which include the normal
  derivative are appropriate boundary conditions on $\partial \Omega$ for all
  tracers. Instead of the inhomogeneous Neumann boundary conditions taking
  the flux interactions, for example, with the atmosphere or sediment into
  account, we apply the homogeneous Neumann boundary conditions without tracer
  fluxes on the boundary.

  Both the advection and the diffusion determine the tracer transport in marine
  water. With a given velocity field $v: \Omega \times [0,1] \rightarrow
  \mathbb{R}^3$, the advection is defined as
  \begin{align}
    \label{eqn:Advection}
    A(x,t) y_i (x,t) &:= \textrm{div} \left( v(x,t) y_i (x,t) \right),
    & x \in \Omega, t &\in [0,1]
  \end{align}
  for $i \in \{1, \ldots, n_y\}$. In ocean circulation modeling, the horizontal
  and vertical direction of the diffusion are usually considered separately
  because of the quite different spatial scales requiring an implicit treatment
  of the vertical part in time integration. Thus, the diffusion operator
  $D = D_h + D_v$ consists of a horizontal and a vertical part defined by
  \begin{align}
    \label{eqn:Diffusion-horizontal}
    D_{h} (x,t) y_i (x,t) &:= -{\rm div}_h \left( \kappa_h (x,t) \nabla_h
                                y_i (x,t) \right)
    & x \in \Omega, t &\in [0,1],\\
    \label{eqn:Diffusion-vertical}
    D_{v} (x,t) y_i (x,t) &:= -\frac{\partial}{\partial z}
            \left(\kappa_{v} (x,t)\frac{\partial y_i}{\partial z}(x,t)\right),
    & x \in \Omega, t &\in [0,1]
  \end{align}
  for $i \in \{1, \ldots, n_y\}$, where $\textrm{div}_h$ and $\nabla_h$ denote
  the horizontal divergence and gradient, $\kappa_h, \kappa_v: \Omega \times
  [0,1] \rightarrow \mathbb{R}$ the diffusion coefficient fields and $z$ the
  vertical coordinate. The diffusion coefficient fields $\kappa_h, \kappa_v$
  and, therefore, the tracer transport are identical for all tracers if the
  molecular diffusion of the tracers is considered to be negligible compared to
  the turbulent mixing, which is a suitable simplification.

  The biogeochemical model includes the biogeochemical processes within the
  ecosystem. In contrast to the biogeochemical model, the marine ecosystem model
  additionally contains the effects of the ocean circulation and, thus,
  represents the whole system \eqref{eqn:Modelequation} to
  \eqref{eqn:PeriodicCondition}. For $i \in \{1, \ldots, n_{y}\}$, the generally
  nonlinear function $q_i: \Omega \times [0,1] \rightarrow \mathbb{R},
  \left( x, t \right) \mapsto q_i \left( x, t, \vec{y}, \vec{u} \right)$
  describes the biogeochemical processes including the coupling to the other
  tracers for the tracer $y_i$. These functions $q_i$ depend firstly on space
  and time due to the variability of the solar radiation and its influence on
  the biogeochemical processes, secondly on the different tracers and thirdly on
  $n_u \in \mathbb{N}$ model parameters (for example growth, loss and mortality
  rates or sinking speeds) summarized in a parameter vector $\vec{u} \in
  \mathbb{R}^{n_u}$. Altogether, these functions form the biogeochemical model
  $\vec{q} = \left( q_i \right)_{i=1}^{n_{y}}$.

  For the marine ecosystem model, our aim is the calculation of a steady annual
  cycle, i.e., a periodic solution of \eqref{eqn:Modelequation} and
  \eqref{eqn:Boundarycondition}, which additionally satisfies
  \begin{align}
   \label{eqn:PeriodicCondition}
    y_i (x, 0) &= y_i (x, 1), & x &\in \Omega,
  \end{align}
  for $i \in \{1, \ldots, n_y\}$. For this purpose, we assume that the operators
  $D$, $A$ and the functions $q_i$ are annually periodic in time.

\subsection{Biogeochemical models}
\label{sec:Model-BiogeochemicalModel}

  We applied a hierarchy of global biogeochemical models with an increasing
  complexity introduced by \citet{KrKhOs10}. \citet{KrOsKh12} and
  \citet{PiwSla16} already applied these biogeochemical models for diverse
  experiments. In the following, we provide a short description of each model
  following the notation by \citet{PiwSla16} and refer to \citet{KrKhOs10} and
  \citet{PiwSla16} for a detailed description of the modeled processes and
  model equations of each biogeochemical model in this hierarchy and, in
  particular, to \citet[][Appendix B]{PiwSla16} for the equations of the
  different tracers. Table \ref{table:Parameter-Modelhierarchy} summarizes the
  model parameters for the different models of the hierarchy.

  \begin{table*}[t]
    \caption{Parameter values for the model hierarchy
             $(\mathbf{u}_{\text{ref}})$ as well as lower $(\vec{b}_\ell)$ and
             upper $(\vec{b}_u)$ bounds for the parameter values used to
             generate the Latin hypercube sample for the whole model hierarchy.}
  \label{table:Parameter-Modelhierarchy}
  \begin{tabular}{l l r c c l}
  \tophline
  Parameter & Description & $\mathbf{u}_{\text{ref}}$ & $\vec{b}_\ell$ & $ \vec{b}_u$ & Unit \\
  \middlehline
  $k_w$ & Attenuation coefficient of water & 0.02 & 0.01 & 0.05 & \unit{m^{-1}} \\
  $k_c$ & Attenuation coefficient of phytoplankton & 0.48 & 0.24 & 0.72 & \unit{(mmol\, P\, m^{-3})^{-1} m^{-1}} \\
  $\mu_P$ & Maximum growth rate & 2.0 & 1.0   & 4.0  & \unit{d^{-1}} \\
  $\mu_Z$ & Maximum grazing rate & 2.0 & 1.0   & 4.0  & \unit{d^{-1}} \\
  $K_N$ & Half saturation constant for \unit{PO_4} uptake & 0.5 & 0.25  & 1.0  & \unit{mmol\, P\, m^{-3}} \\
  $K_P$ & Half saturation constant for grazing & 0.088 & 0.044 & 0.176 & \unit{mmol\, P\, m^{-3}} \\
  $K_I$ & Light intensity compensation & 30.0 & 15.0 & 60.0 & \unit{W\, m^{-2}} \\
  $\sigma_Z$ & Fraction of production remaining in zooplankton & 0.75 & 0.05 & 0.95 & \unit{1} \\
  $\sigma_\text{DOP}$ & Fraction of phytoplankton and zooplankton losses assigned to DOP & 0.67 & 0.05 & 0.95 & \unit{1} \\
  $\lambda_P$ & Linear phytoplankton loss rate & 0.04 & 0.02 & 0.08 & \unit{d^{-1}} \\
  $\kappa_P$ & Quadratic phytoplankton loss rate & 4.0 & 2.0 & 6.0 & \unit{(mmol\, P\, m^{-3})^{-1} d^{-1}} \\
  $\lambda_Z$ & Linear zooplankton loss rate & 0.03 & 0.015 & 0.045 & \unit{d^{-1}} \\
  $\kappa_Z$ & Quadratic zooplankton loss rate & 3.2 & 1.6 & 4.8 & \unit{(mmol\, P\, m^{-3})^{-1} d^{-1}} \\
  $\lambda'_P$ & Phytoplankton mortality rate & 0.01 & 0.005 & 0.015 & \unit{d^{-1}} \\
  $\lambda'_Z$ & Zooplankton mortality rate & 0.01 & 0.005 & 0.015 & \unit{d^{-1}} \\
  $\lambda'_D$ & Degradation rate & 0.05 & 0.025 & 0.1 & \unit{d^{-1}} \\
  $\lambda'_\text{DOP}$ & Decay rate & 0.5 & 0.25 & 1.0 & \unit{yr^{-1}} \\
  $b$ & Implicit representation of sinking speed & 0.858 & 0.7 & 1.5 & \unit{1}  \\
  $a_D$ & Increase of sinking speed with depth & 0.058 & 0.029 & 0.087 & \unit{d^{-1}} \\
  $b_D$ & Initial sinking speed & 0.0 & 0.0 & 0.0 & \unit{m\,d^{-1}} \\
  \bottomhline
  \end{tabular}
  \end{table*}

  For all biogeochemical models considered here, the light intensity influences
  the biogeochemical processes, especially the biological production. The light
  limitation function $I: \Omega \times [0, 1] \rightarrow \mathbb{R}_{\geq 0}$
  depends on the insolation computed on the fly using the astronomical formula
  of \citet{PalPla76}, and takes into account the ice cover as well as the
  exponential attenuation of water and (if included in the model) phytoplankton.
  Based on light intensity, the ocean $\Omega$ is divided into a euphotic (sun
  lit) zone of about \SI{100}{\metre} and an aphotic zone below. The euphotic
  zone, in particular, exhibits a fast and dynamic turnover of phosphorus, for
  example through photosynthesis, grazing or mortality. One part of the
  biological production sinks as particulate matter from the euphotic zone to
  depth, where it is remineralized according to the empirical power law
  relationship discovered by \citet{MKKB87}.

  The simplest model of the hierarchy takes phosphate (\unit{PO_4}) only as
  inorganic nutrient into account \citep{KrKhOs10}. We denote this model as N
  model (N for nutrients) and $\vec{y} = \left( \vec{y}_{\text{N}}
  \right)$. The phytoplankton production
  \begin{align}
    \label{eqn:Phytoplankton}
    f_P: \Omega \times [0,1] \rightarrow \mathbb{R},
      f_P (x, t) &= \mu_P y_P^* \frac{I(x,t)}{K_I + I(x,t)}
                    \frac{\vec{y}_N (x,t)}{K_N + \vec{y}_N (x,t)}
  \end{align}
  depends on the light intensity and is limited by using a half saturation
  function, a maximum production rate parameter $\mu_P$ and a prescribed
  concentration of phytoplankton $y_P^* = 0.0028$ \unit{mmol\, P\, m^{-3}}. This
  model contains $n_u = 5$ model parameters $\vec{u} = \left( k_w, \mu_P, K_N,
  K_I, b \right)$ describing the internal processes, and resembles the model of
  \citet{BacMai90}.

  The second model of this hierarchy includes dissolved organic phosphorus
  (\unit{DOP}) in addition to nutrients (\unit{N})
  \citep[cf.][]{KrKhOs10, BacMai91, PaFoBo05}. Hence, we denote this model as
  N-DOP model with the tracers $\vec{y} = \left( \vec{y}_{\text{N}},
  \vec{y}_{\text{DOP}} \right)$. The $n_u = 7$ model parameters $\vec{u} =
  \left( k_w, \mu_P, K_N, K_I, \sigma_{\text{DOP}}, \lambda'_{\text{DOP}},
  b \right)$ control the internal processes using the unchanged phytoplankton
  production \eqref{eqn:Phytoplankton}.

  Resolving explicitly phytoplankton (\unit{P}) increases the model complexity
  receiving the NP-DOP model, i.e., $\vec{y} = \left(
  \vec{y}_{\text{N}}, \vec{y}_{\text{P}}, \vec{y}_{\text{DOP}} \right)$
  \citep{KrKhOs10}. Instead of using $y_P^*$ in \eqref{eqn:Phytoplankton}, the
  phytoplankton production (explicitly) includes the phytoplankton concentration
  $\vec{y}_P$. In addition, the phytoplankton concentration affects the light
  intensity. Moreover, the model contains a term for zooplankton grazing
  \begin{align}
    \label{eqn:Zooplankton}
    f_Z: \Omega \times [0,1] \rightarrow \mathbb{R},
    f_Z (x,t) &= \mu_Z y_Z^* \frac{\vec{y}_P(x,t)^2}{K_P^2 + \vec{y}_P(x,t)^2}
  \end{align}
  using the implicitly prespecified zooplankton concentration $y_Z^* = 0.01$
  \unit{mmol\, P\, m^{-3}}. Overall, this model includes the $n_u = 13$ model
  parameters $\vec{u} = \left( k_w, k_c, \mu_P, \mu_Z, K_N, K_P, K_I,
  \sigma_{\text{DOP}}, \lambda_{P}, \kappa_P, \lambda'_{P},
  \lambda'_{\text{DOP}}, b \right)$.

  Adding zooplankton (\unit{Z}) increases again the complexity. We denote this
  model as NPZ-DOP model with the tracers $\vec{y} = \left( \vec{y}_{\text{N}},
  \vec{y}_{\text{P}}, \vec{y}_{\text{Z}}, \vec{y}_{\text{DOP}} \right)$
  \citep{KrKhOs10}. While the phytoplankton production is the same as for the
  NP-DOP model, the zooplankton grazing \eqref{eqn:Zooplankton} includes
  (explicitly) the zooplankton concentration $\vec{y}_Z$ instead of using
  $y_Z^*$. The $n_u = 16$ model parameters are $\vec{u} = \left( k_w, k_c,
  \mu_P, \mu_Z, K_N, K_P, K_I, \sigma_{Z}, \sigma_{\text{DOP}}, \lambda_{P},
  \lambda_{Z}, \kappa_{Z}, \lambda'_{P}, \lambda'_{Z}, \lambda'_{\text{DOP}},
  b \right)$.

  The last model of the hierarchy, the NPZD-DOP model, is similar to the model
  introduced by \citet{SOGES05} and explicitly resolves detritus (\unit{D}),
  i.e., $\vec{y} = \left( \vec{y}_{\text{N}} , \vec{y}_{\text{P}},
  \vec{y}_{\text{Z}}, \vec{y}_{\text{D}}, \vec{y}_{\text{DOP}} \right)$. Both
  the phytoplankton production and the zooplankton grazing are unchanged. The
  model contains the model parameters $\vec{u} = \left( k_w, k_c,
  \mu_P, \mu_Z, K_N, K_P, K_I, \sigma_{Z}, \sigma_{\text{DOP}}, \lambda_{P},
  \lambda_{Z}, \kappa_{Z}, \lambda'_{P}, \lambda'_{Z}, \right.$ $\left.
  \lambda'_{D}, \lambda'_{\text{DOP}}, a_D, b_D \right)$, i.e., $n_u = 18$
  \citep[cf.][]{KrKhOs10}.

  In addition to the model hierarchy, we applied the model of
  \citet{DuSoScSt05} used for the MIT general circulation model
  biogeochemistry tutorial \citep[cf.][]{MAHPH97, SSDPKJ05}. This model includes
  phosphate (\unit{PO_4}) and dissolved organic phosphorus (\unit{DOP}) and
  resembles the N-DOP model. We denote this model as MITgcm-PO4-DOP model.
  The $n_u = 7$ model parameters are shown in Table
  \ref{table:Parameter-MITgcm-PO4-DOP} and correspond to the model parameters of
  the N-DOP model with different parameter names.

  \begin{table*}[tb]
    \caption{Parameter values $(\vec{u}_\text{ref})$ used for the MITgcm-PO4-DOP
             model and the symbol used by \cite{DuSoScSt05} as well as lower
             $(\vec{b}_\ell)$ and upper $(\vec{b}_u)$ bound for the parameter
             values used to generate the Latin hypercube sample.}
    \label{table:Parameter-MITgcm-PO4-DOP}
    \begin{tabular}{l l c c c l}
      \tophline
      Parameter & Description & $\vec{u}_\text{ref}$ &  $\vec{b}_\ell$ & $ \vec{b}_u$ & Unit \\
      \middlehline
      $\kappa_{\text{remin}}$ & Decay rate & 0.5 & 0.25 & 1.0 & \unit{yr^{-1}} \\
      $\alpha$ & Maximum growth rate & 2.0 & 1.0 & 4.0 & \unit{mmol\, P\, m^{-3}} \\
      $f_{\text{DOP}}$ & Fraction of phytoplankton losses assigned to DOP & 0.67 & 0.05 & 0.095 & \unit{1} \\
      $\kappa_{\text{PO}_4}$ & Half saturation constant for \unit{PO_4} uptake & 0.5 & 0.25 & 1.0 & \unit{mmol\, P\, m^{-3}} \\
      $\kappa_{\text{I}}$ & Light intensity compensation & 30.0 & 15.0 & 60.0 & \unit{W\, m^{-2}} \\
      $k$ & Attenuation coefficient of water & 0.02 & 0.01 & 0.05 & \unit{m^{-1}} \\
      $a_{\text{remin}}$ & Implicit representation of sinking speed & 0.858 & 0.7 & 1.5 & \unit{1} \\
      \bottomhline
    \end{tabular}
  \end{table*}

\subsection{Transport matrix method}
\label{sec:Model-TMM}

  The transport matrix method (TMM) introduced by \cite{KhViCa05} is an
  efficient offline method for the simulation of tracer transports
  \citep[cf.][]{Kha07}. For offline models, the ocean circulation takes the
  velocity field $v$ and the diffusion fields $\kappa_h$, $\kappa_v$ into
  account that can be pre-computed by some ocean circulation model running into
  a steady annual cycle. The TMM takes advantage of the fact that the
  application of the two operators $D$ and $A$ on a spatially discretized tracer
  vector is linear. Therefore, the discretized advection-diffusion equation can
  be written as a linear matrix equation. Hence, the TMM approximates the ocean
  circulation by matrices rather than to implement a discretization scheme for
  diffusion and advection. Instead of storing the fields $v, \kappa_h$ and
  $\kappa_v$, the TMM computes and stores uniquely the matrices which represent
  the application of the discretized operators on a discrete tracer vector. This
  approach ensures that the matrices contain the transport of all parameterized
  processes represented in the ocean circulation model \citep{KhViCa05}.

  The TMM reduces a time step of the marine ecosystem model to two
  matrix-vector-multiplications and one evaluation of the biogeochemical model.
  We assume that the grid with $n_x \in \mathbb{N}$ grid points $\left( x_k
  \right)_{k=1}^{n_x}$ is a spatial discretization of the domain
  $\Omega$  (i.e., the ocean) and the time steps $t_0, \ldots, t_{n_{t}} \in
  [0,1]$, $n_t \in \mathbb{N}$, specified by
  \begin{align*}
    t_j &:= j \Delta t, & j &= 0, \ldots, n_t, & \Delta t &:= \frac{1}{n_t},
  \end{align*}
  define an equidistant grid of the time interval $[0,1]$  (i.e., one model
  year). For the time instant $t_j$, $j \in \{0, \ldots, n_t - 1\}$, the vector
  \begin{align*}
    \vec{y}_{ji} &\approx \left( y_{i} \left( t_{j}, x_{k} \right)
                         \right)_{k=1}^{n_x} \in \mathbb{R}^{n_x}
  \end{align*}
  contains the numerical approximation of the spatially discrete tracer $y_i$,
  $i \in \{1, \ldots, n_y\}$, at the fixed time instant $t_j$ and $\vec{y}_{j}
  := \left( \vec{y}_{ji} \right)_{i=1}^{n_y} \in \mathbb{R}^{n_y n_x}$ includes
  the numerical approximation of all tracers at time instant $t_j$ using a
  reasonable concatenation. Analogously,
  \begin{align*}
    \vec{q}_{ji} &\approx \left( q_i \left( x_k, t_j, \vec{y_j}, \vec{u} \right)
                     \right)_{k=1}^{n_x} \in \mathbb{R}^{n_x}
  \end{align*}
  denotes the spatially discretized biogeochemical term of tracer $y_i$, $i \in
  \{1, \ldots, n_y\}$, at time instant $t_j$, $j \in \{0, \ldots, n_t - 1\}$,
  and $\vec{q}_j := \left( \vec{q}_{ji} \right)_{i=1}^{n_y}$ the spatially
  discretized biogeochemical term of all tracers at time instant $t_j$. Applying
  a semi-implicit Euler scheme, where the advection and the horizontal
  diffusion are discretized explicitly and the vertical diffusion implicitly,
  the discretization of \eqref{eqn:Modelequation} results in a time-stepping
  \begin{align*}
    \vec{y}_{j+1} &=  \left( \mathbf{I} + \Delta t \mathbf{A}_j
                            + \Delta t \mathbf{D}_j^h \right) \vec{y}_j
                     + \Delta t \mathbf{D}_j^v \vec{y}_{j+1}
                     + \Delta t \vec{q}_j \left( \vec{y}_j, \vec{u} \right),
        & j &= 0, \ldots, n_t -1
  \end{align*}
  with the identity matrix $\mathbf{I} \in \mathbb{R}^{n_x \times n_x}$ and the
  spatially discretized counterparts $\mathbf{A}_j, \mathbf{D}_j^h$ and
  $\mathbf{D}_j^v$ of the operators $A, D_h$ and $D_v$ at time instant $t_j$,
  $j \in \{0, \ldots, n_t - 1\}$. Defining the explicit and implicit transport
  matrices
  \begin{align*}
    \mathbf{T}_{j}^{\text{exp}} &:= \mathbf{I} + \Delta t \mathbf{A}_j
                                    + \Delta t \mathbf{D}_j^h \in \mathbb{R}^{n_x \times n_x}, \\
    \mathbf{T}_{j}^{\text{imp}} &:= \left( \mathbf{I} - \Delta t \mathbf{D}_j^v
                                    \right)^{-1} \in \mathbb{R}^{n_x \times n_x}
  \end{align*}
  for each time instant $t_j$, $j \in \{0, \ldots, n_t - 1\}$, a time step of
  the marine ecosystem model using the TMM is specified by
  \begin{align}
    \label{eqn:TMM}
    \vec{y}_{j+1} &= \mathbf{T}_{j}^{\text{imp}}
                     \left( \mathbf{T}_{j}^{\text{exp}} \vec{y}_j
                        + \Delta t \vec{q}_j \left( \vec{y}_j, \vec{u} \right)
                        \right)
                 =: \varphi_j \left( \vec{y}_j, \vec{u} \right),
        & j &= 0, \ldots, n_t - 1.
  \end{align}

  The transport matrices are sparse and represent the monthly averaged tracer
  transport respectively. Due to the calculation of the matrices with a
  grid-point based ocean circulation model, the explicit ones are sparse.
  Moreover, the implicit ones (i.e., the inverse of the discretization matrices)
  are sparse because they contain only the vertical part of the diffusion.
  Storing the transport matrices for all time-steps in a year is practically
  impossible. Therefore, monthly averaged matrices are stored. As a consequence,
  the matrices are interpolated linearly to compute an approximation for any
  time instant $t_j$, $j = 0, \ldots, n_t - 1$. Assuming annual periodicity of
  the ocean circulation, twelve pairs of pre-computed transport matrices
  approximate the ocean circulation. \citet{KhViCa05} presented the details of
  the transport matrix computation.

  We have used twelve explicit and twelve implicit transport matrices
  representing the monthly averaged tracer transport. These matrices are derived
  from the MIT ocean model \citep{MAHPH97} using a global configuration with a
  latitudinal and longitudinal resolution of \ang{2.8125} with \num{15} vertical
  layers.

\subsection{Computation of steady annual cycles}
\label{sec:SteadyAnnualCycle}

  For the marine ecosystem model, the steady annual cycle is a fixed-point of
  time integration over one model year. The steady annual cycle defined as a
  periodic solution of the system \eqref{eqn:Modelequation} to
  \eqref{eqn:PeriodicCondition} with a period length of one model year fulfills
  \begin{align*}
    \vec{y}_{n_t} &= \vec{y}_0
  \end{align*}
  in the fully discrete setting applying the above iteration \eqref{eqn:TMM}.
  The nonlinear mapping
  \begin{align*}
    \Phi &:= \varphi_{n_t -1} \circ \ldots \circ \varphi_{0}
  \end{align*}
  with $\varphi_j$ defined in \eqref{eqn:TMM} corresponds to the time
  integration of \eqref{eqn:TMM} over one model year. In particular, a
  steady annual cycle is a fixed-point of the mapping $\Phi$. Starting from an
  arbitrary vector $\vec{y}^{0} \in \mathbb{R}^{n_y n_x}$, a classical
  fixed-point iteration takes the form
  \begin{align}
    \label{eqn:Spin-upIteration}
    \vec{y}^{\ell + 1} &= \Phi \left( \vec{y}^{\ell}, \vec{u} \right),
    & \ell = 0, 1, \ldots.
  \end{align}
  The vector $\vec{y}^{\ell} := \vec{y}_{\left( \ell - 1 \right) n_t} \in
  \mathbb{R}^{n_y n_x}$ contains the tracer concentrations at the first time
  instant of model year $\ell \in \mathbb{N}$ if we interpret the fixed-point
  iteration as pseudo-time stepping or \emph{spin-up}. If $\Phi$ is a
  contraction mapping fulfilling
  \begin{align*}
    \left\| \Phi \left(\vec{x}, \vec{u} \right) - \Phi \left( \vec{z}, \vec{u}
    \right) \right\| \leq L \left\| \vec{x} - \vec{z} \right\|
  \end{align*}
  for all $x, z \in \mathbb{R}^{n_y n_x}$ with $L \in [0,1[$ in some norm,
  the Banach fixed-point theorem \citep[cf.][]{Ban22,DahReu08} ensures the
  convergence of the spin-up towards a unique fixed-point for all initial
  tracer concentrations $\vec{y}^0 \in \mathbb{R}^{n_y n_x}$. This result is
  also valid for weaker assumptions \citep[cf.][]{Cir74}. This approach of
  computing a fixed-point yields a robust method. However, the convergence
  behavior is only linear, and the estimation of $L = \max_{\vec{z} \in
  \mathbb{R}^{n_y n_x}} \left\| \Phi' \left( \vec{z}, \vec{u} \right) \right\|$
  is difficult due to the involved Jacobian $\vec{q}'_j \left( \vec{y}_j,
  \vec{u} \right)$ of the nonlinear biogeochemical model for the current time
  instant $t_j$, $j \in \{0, \ldots, n_t - 1 \}$ . In order to reach a steady
  annual cycle the spin-up ordinarily requires several thousand applications of
  the mapping $\Phi$ (i.e., model years) \citep[cf.][]{BeDiWu08}. In order to
  test the numerical convergence of the iteration \eqref{eqn:Spin-upIteration},
  we measured the periodicity of the steady annual cycle by the difference
  between two consecutive iterates defined by
  \begin{align}
    \label{eqn:StoppingCriterion}
    \varepsilon_{\ell} := \left\| \vec{y}^{\ell} - \vec{y}^{\ell - 1} \right\|
  \end{align}
  for iteration (i.e., model year) $\ell \in \mathbb{N}$.

\subsection{Norms}
\label{sec:Norms}

  We used various norms to quantify the difference between different tracer
  concentrations. For $\vec{z} \in \mathbb{R}^{n_{y} n_{x}}$ indexed as
  $\vec{z} = \left( \left( z_{ik} \right)_{k=1}^{n_{x}} \right)_{i=1}^{n_{y}}$
  in accordance with the tracer indexing in Sect. \ref{sec:Model-TMM}, we
  defined a weighted Euclidean norm
  \begin{align*}
       \left\| \vec{z} \right\|_{2, w} &:= \left( \sum_{i=1}^{n_{y}} 
          \sum_{k=1}^{n_{x}} w_k z_{ik}^2 \right)^{\frac{1}{2}}
  \end{align*}
  with weights $w_k \in \mathbb{R}_{> 0}$ for $k \in \left\{1, \ldots, n_{x}
  \right\}$. A special case of this weighted Euclidean norm $\left\| \cdot
  \right\|_{2, w}$ is the Euclidean norm $\left\| \cdot \right\|_2$ if $w_k = 1$
  holds for all $k \in \left\{1, \ldots, n_{x} \right\}$. We introduced the norm
  $\left\| \cdot \right\|_{2, V}$ as a weigthed Euclidean norm $\left\| \cdot
  \right\|_{2, w}$ with the weights $w_k = \left| V_k \right|$, $k \in \left\{1,
  \ldots, n_{x} \right\}$, where $\left| V_k \right|$ denotes the box volume
  corresponding to the grid point $x_k$. This norm $\left\| \cdot
  \right\|_{2,V}$ is the discretized counterpart of the $\left( L^2 (\Omega)
  \right)^{n_{y}}$ norm. For all weight vectors $\vec{w} \in
  \mathbb{R}_{>0}^{n_{x}}$ and any vector $\vec{z} \in
  \mathbb{R}^{n_{y} n_{x}}$, particularly, holds
  \begin{align*}
      \min_{1 \leq k \leq n_{x}} \sqrt{w_k} \left\| \vec{z} \right\|_2
          &\leq \left\| \vec{z} \right\|_{2, w}
           \leq \max_{1 \leq k \leq n_{x}} \sqrt{w_k} \left\| \vec{z} \right\|_2,
  \end{align*}
  i.e., the Euclidean norm $\left\| \cdot \right\|_2$ and the weighted Euclidean
  norm $\left\| \cdot \right\|_{2,w}$ are equivalent in the mathematical sense
  for all positive weights. Furthermore, we used a weighted Euclidean norm
  $\left\| \cdot \right\|_{2, w, T}$ for the whole trajectory of all tracers
  over one model year. For a vector $\vec{z} \in \mathbb{R}^{n_{t} n_{y} n_{x}}$
  indexed as $\vec{z} = \left( \left( \left( z_{jik} \right)_{k=1}^{n_{x}}
  \right)_{i=1}^{n_{y}} \right)_{j=0}^{n_{t}-1}$ and a weight vector $\vec{w}
  \in \mathbb{R}^{n_{x}}_{>0}$, we defined the norm
  \begin{align*}
      \left\| \vec{z} \right\|_{2, w, T} &:= \left( \sum_{i=1}^{n_{y}}
        \sum_{j=0}^{n_{t}-1} \Delta t \sum_{k=1}^{n_{x}} w_k z_{jik}^2
        \right)^{\frac{1}{2}}.
  \end{align*}
  We denote in analogy to the weighted Euclidean norm by $\left\| \cdot \right
  \|_{2, T}$ the norm $\left\| \cdot \right\|_{2,w,T}$ with weights $w_k = 1$,
  $k \in \{1, \ldots, n_x\}$, and by $\left\| \cdot \right\|_{2,V,T}$ the
  weighted Euclidean norm $\left\| \cdot \right\|_{2,w,T}$ where the weights are
  set to $w_k := \left| V_k \right|$, $k \in \left\{1, \ldots, n_{x} \right\}$.

\subsection{Influence of model to data misfit}
\label{sec:Costfunction}

  We evaluated the steady annual cycle by measuring the difference between the
  calculated tracer concentrations and observations from the world ocean
  database \citep{WOD13, Rei19}. The approximation of the observations by the
  calculated steady annual cycles were comparable when using the following norm
  for different time steps. The norm is based on norm $\left\| \cdot
  \right\|_{2,T}$ taking into account only points in space and time for which
  there are observations available. Thereby, we applied only the quality
  controlled datasets. In order to take the number of measurements at the same
  point in space and time into account, we applied the world ocean database
  instead of the world ocean atlas \citep{Gar14}. There are available both
  points in space and time with many measurements and points for which there are
  no measurements. Hence, we skipped the discrete points in space and time
  without measurements in the norm. If many measurements exist at a discrete
  point in space and time, the norm, by contrast, involved each measurement
  individually.

  We denoted by $n_{m_{ijk}} \in \mathbb{N}_0$ the number of measurements for
  the given tracer $\vec{y}_i$, $i \in \{1, \ldots, n_y\}$, the point in time
  $t_j$, $j \in \{0, \ldots, n_{t}-1\}$ and the grid point $x_k$, $k \in \{1,
  \ldots, n_x\}$. We combined all quality controlled datasets of the world ocean
  database in the vector $\vec{y}_{\text{data}} \in \mathbb{R}^{n_m}$ indexed as
  $\vec{y}_{\text{data}} = \left( \left( \left( \left( y_{\text{data}, jikl}
  \right)_{l = 1}^{n_{m_{ijk}}} \right)_{k=1}^{n_x} \right)_{i=1}^{n_y}
  \right)_{j=0}^{n_{t}-1}$ with $n_m := \sum_{i=1}^{n_y} \sum_{j=0}^{n_{t}-1}
  \sum_{k=1}^{n_x} n_{m_{ijk}}$. We defined the discrete cost function of
  ordinary least squares (OLS) \citep{SebWil03} as
  \begin{align}
   \label{eqn:Costfunction}
   J_{\text{OLS}} \left(\vec{z}\right)
              &:= \left\| \vec{z} - \vec{y}_\text{data} \right\|_{2, T,
                   \text{OLS}}^2
               := \sum_{i=1}^{n_y} \sum_{j=0}^{n_{t}-1} \sum_{k=1}^{n_x}
                  \sum_{l = 1}^{n_{m_{ijk}}} \left( z_{jik} - y_{\text{data},
                jikl} \right)^2
  \end{align}
  for a vector $\vec{z} \in \mathbb{R}^{n_t n_y n_x}$ indexed as $\vec{z} =
  \left( \left( \left( z_{jik} \right)_{k=1}^{n_{x}} \right)_{i=1}^{n_{y}}
  \right)_{j=0}^{n_{t}-1}$.

\section{Larger time steps}
\label{sec:Timesteps}

  The increasing of the time step $\Delta t$ leads to a shortening of the runtime
of the spin-up. An increase of the time step reduces the computational effort
because the spin-up needs several thousand model years to reach the steady
annual cycle \citep[cf.][]{BeDiWu08}. However, the dynamics in the ocean
circulation limit the time step in marine ecosystem models to a few hours when
the dynamical and tracer equations are integrated simultaneously \citep{Kha07}.
Consequently, the computational effort is enormous for marine ecosystem models
involving biogeochemical tracers to compute a steady annual cycle.

The transport matrices can be adjusted to the larger time step by simple matrix
operations. When the transport matrices are generated, the time step of the
underlying ocean circulation model directly enters each transport matrix.
Nevertheless, \citet{Kha07} described a possibility of the TMM to utilize larger
time steps with minimal loss of accuracy. We generated transport matrices
\begin{align}
  \label{eqn:TMMTimestepsExp}
  \mathbf{T}_{j, m}^{\text{exp}} &:= \mathbf{I} + m \left(
                            \mathbf{T}_{j}^{\text{exp}} - \mathbf{I} \right) \\
  \label{eqn:TMMTimestepsImp}
  \mathbf{T}_{j, m}^{\text{imp}} &:= \left( \mathbf{T}_{j}^{\text{imp}} \right)^m
\end{align}
for every time instant $t_j$, $j \in \{0, \ldots, n_{t}-1\}$, whose effective
time step is by a factor $m \in \mathbb{N}$ larger compared to the underlying
ocean circulation model. The explicit transport matrix
$\mathbf{T}_{j, m}^{\text{exp}}$ is, thereby, the exact representation of the
rougher time step while the implicit transport matrix
$\mathbf{T}_{j, m}^{\text{imp}}$ is an approximation only which is
asymptotically correct \citep{PiwSla16a}. The transport matrices for rougher
time steps maintain even their sparsity \citep{Kha07}.

Assuming 360 days per year, the temporal resolution of the ocean circulation
model used for the computation of the transport matrices
$\mathbf{T}_{j}^{\text{exp}}, \mathbf{T}_{j}^{\text{imp}}$, $j \in \{0 \ldots,
n_{t} - 1\}$, corresponds to \SI{3}{\hour}. Accordingly, the number of time
steps per year is $n_{t} = 2880$. We identified this temporal resolution with
time step \SI{1}{\Timestep}. Additionally, the transport matrices
$\mathbf{T}_{j, 1}^{\text{exp}}, \mathbf{T}_{j, 1}^{\text{imp}}$, $j \in \{0,
\ldots, n_{t} - 1\}$, belong to this time step \SI{1}{\Timestep}.

\begin{table*}
  \caption{Temporal resolution of the TMM using different time steps.}
  \label{table:Timesteps}
  \begin{tabular}{c c c}
    \tophline
    Time step & $n_{t}$ & Corresponding time steps \\
    \middlehline
    \SI{ 1}{\Timestep} & \num{2880} &   \SI{3}{\hour} \\
    \SI{ 2}{\Timestep} & \num{1440} &   \SI{6}{\hour} \\
    \SI{ 4}{\Timestep} &  \num{720} &  \SI{12}{\hour} \\
    \SI{ 8}{\Timestep} &  \num{360} &  \SI{24}{\hour} \\
    \SI{16}{\Timestep} &  \num{180} &  \SI{48}{\hour} \\
    \SI{32}{\Timestep} &   \num{90} &  \SI{96}{\hour} \\
    \SI{64}{\Timestep} &   \num{45} & \SI{192}{\hour} \\
    \bottomhline
  \end{tabular}
\end{table*}

We have studied the time steps listed in Table \ref{table:Timesteps}. Time step
\SI{2}{\Timestep} conformed, for instance, to a doubling of the effective time
step in \eqref{eqn:TMMTimestepsExp} and \eqref{eqn:TMMTimestepsImp}. It is
obvious that the use of larger time steps reduces the number of steps per year.
In order to distinguish the tracer concentrations calculated with different time
steps, we defined the vector $\vec{y}^{\ell, t}$ as the vector $\vec{y}^{\ell}$
computed with the time step $t$ \unit{dt} for $\ell \in \mathbb{N}_0$ and
$t \in \{ 1, 2, 4, 8, 16, 32, 64\}$. More importantly, the transport matrices
$\mathbf{T}_{j, t}^{\text{exp}}, \mathbf{T}_{j, t}^{\text{imp}}$, $j \in \{0,
\ldots, n_{t} - 1\}$, replace here the transport matrices
$\mathbf{T}_{j}^{\text{exp}}, \mathbf{T}_{j}^{\text{imp}}$ in \eqref{eqn:TMM}.

\section{Results}
\label{sec:Results}

  In this section, we are presenting the results obtained for the biogeochemical
models introduced in Sect. \ref{sec:Model-BiogeochemicalModel}. We investigated
the accuracy of the calculated steady annual cycles using larger time steps (cf.
Sect. \ref{sec:Timesteps}).

\subsection{Experimental setup}
\label{sec:ExperimentalSetup}

  For the computation of a steady annual cycle, we applied the marine ecosystem
  toolkit for optimization and simulation in 3D (Metos3D), a framework for the
  offline simulation of marine ecosystem models developed by \citet{PiwSla16,
  PiwSla16a}. Here, we computed a spin-up of over \num{10000} model years
  starting with constant global mean tracer concentrations. Except for
  \unit{PO_4}, we initialized each tracer (i.e., the tracer \unit{DOP},
  \unit{P}, \unit{Z} and \unit{D}) with a global mean tracer concentration of
  \SI{0.0001}{\milli\mol\Phosphat\per\cubic\meter} in each simulation.
  \unit{PO_4}, i.e., the nutrient tracer \unit{N}, was initialized with a global 
  mean tracer concentration of \SI{2.17}{\milli\mol\Phosphat\per\cubic\meter}.
  We saved the tracer concentration of all the \num{50}th model years during the
  spin-up.
  
  We computed steady annual cycles for different parameter vectors. First, we
  are showing the results for the parameter vectors shown in Tables
  \ref{table:Parameter-Modelhierarchy} and \ref{table:Parameter-MITgcm-PO4-DOP}
  for each biogeochemical model respectively. Next, we applied a Latin
  hypercube \citep[cf.][]{McBeCo79} sample of \num{100} parameter vectors for
  all model parameters within the bounds of Table
  \ref{table:Parameter-Modelhierarchy}. We created these parameter vectors by
  the \texttt{lhs} routine of \citet{PyDOE17}. We restricted the parameters of
  the created Latin hypercube sample to the required model parameters for each
  model (see Table \ref{table:Parameter-Modelhierarchy}). In order to use these
  Latin hypercube samples for the MITgcm-PO4-DOP model, we identified the
  restricted parameters $\begin{pmatrix} \lambda'_{\text{DOP}}, \mu_P,
  \sigma_{\text{DOP}}, K_N, K_I, k_w, b \end{pmatrix}$ of the N-DOP model in
  this context with the parameters $\begin{pmatrix} \kappa_{\text{remin}},
  \alpha, f_{\text{DOP}}, \kappa_{\text{PO}_4}, \kappa_{\text{I}}, k,
  a_{\text{remin}} \end{pmatrix}$ of the MITgcm-PO4-DOP model (see Table
  \ref{table:Parameter-MITgcm-PO4-DOP}).
  
  We compared the calculation of the steady annual cycle using a larger time
  step with a reference solution, namely the result obtained by a spin-up using
  Metos3D with time step \SI{1}{\Timestep}. For this comparison, we used
  several criteria, such as the norm \eqref{eqn:StoppingCriterion}, the cost
  function \eqref{eqn:Costfunction} and the required model years to reach a
  tolerance of $10^{-4}$ in \eqref{eqn:StoppingCriterion} during the spin-up.
  We measured, in particular, the accuracy of the results of the spin-ups by the
  relative differences
  \begin{align}
    \label{eqn:relativeError}
    \frac{\left\| \vec{y}^{\ell, t} - \vec{y}^{10000, 1} \right\|}
         {\left\| \vec{y}^{10000, 1} \right\|}
         \qquad \text{for} \nobreakspace \ell \in \{50, 100, \ldots, 10000\}, 
                           \nobreakspace t \in \{2, 4, 8, 16, 32, 64\}, \\
    \label{eqn:relativeErrorSurface}
    \frac{\left| \vec{y}^{\ell, t} - \vec{y}^{10000, 1} \right|}
         {\left\| \vec{y}^{10000, 1} \right\|}
         \qquad \text{for} \nobreakspace \ell \in \{50, 100, \ldots, 10000\}, 
                           \nobreakspace t \in \{2, 4, 8, 16, 32, 64\}.
  \end{align}
  We call this quantity \eqref{eqn:relativeError} the \emph{(relative) error}
  of the respective result $\vec{y}^{\ell, t}$.

\subsection{Larger time steps}
\label{sec:Results-LargerTimeSteps}
  
  \begin{figure*}[p]
    \centering
    \subfloat[N model using parameter vector $\mathbf{u}_{\text{N}}$
             \eqref{eqn:ParameterVector_N}. \label{fig:Spinup-N}]{\includegraphics{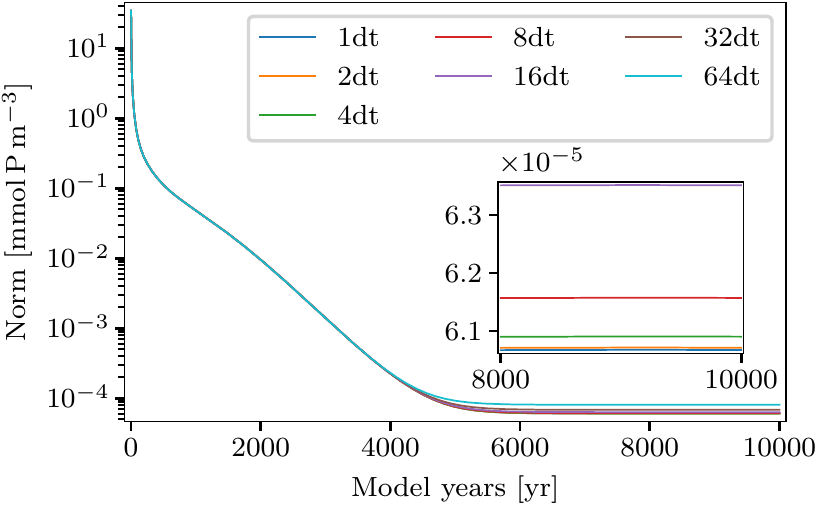}}
    \qquad
    \subfloat[N-DOP model using parameter vector $\mathbf{u}_{\text{N-DOP}}$
             \eqref{eqn:ParameterVector_N-DOP}. \label{fig:Spinup-N-DOP}]{\includegraphics{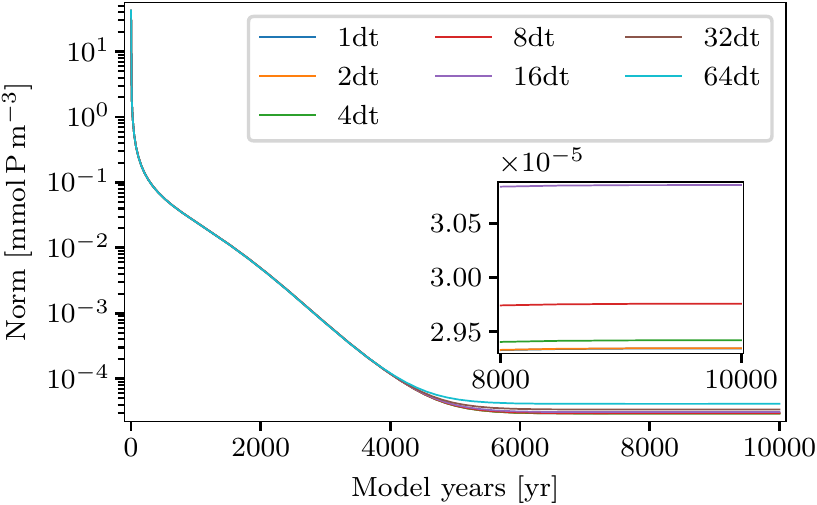}}
    \qquad
    \subfloat[NP-DOP model using parameter vector $\mathbf{u}_{\text{NP-DOP}}$
             \eqref{eqn:ParameterVector_NP-DOP}. \label{fig:Spinup-NP-DOP}]{\includegraphics{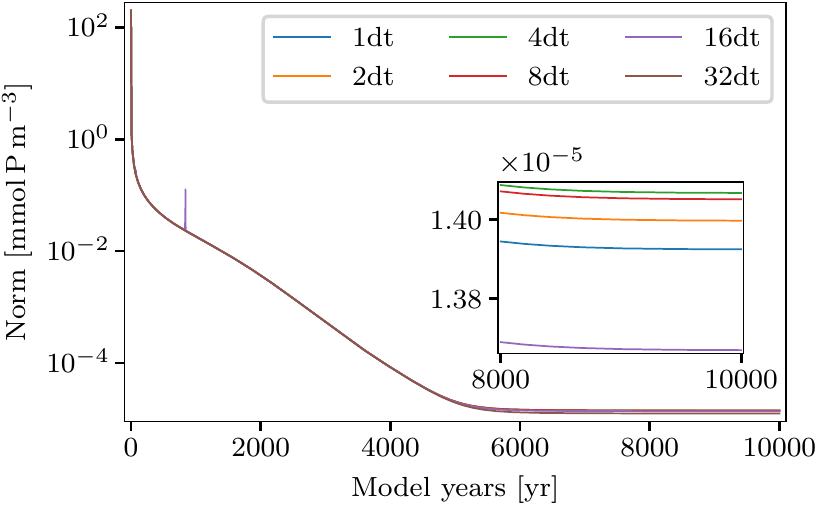}}
    \qquad
    \subfloat[NPZ-DOP model using parameter vector $\mathbf{u}_{\text{NPZ-DOP}}$
             \eqref{eqn:ParameterVector_NPZ-DOP}. \label{fig:Spinup-NPZ-DOP}]{\includegraphics{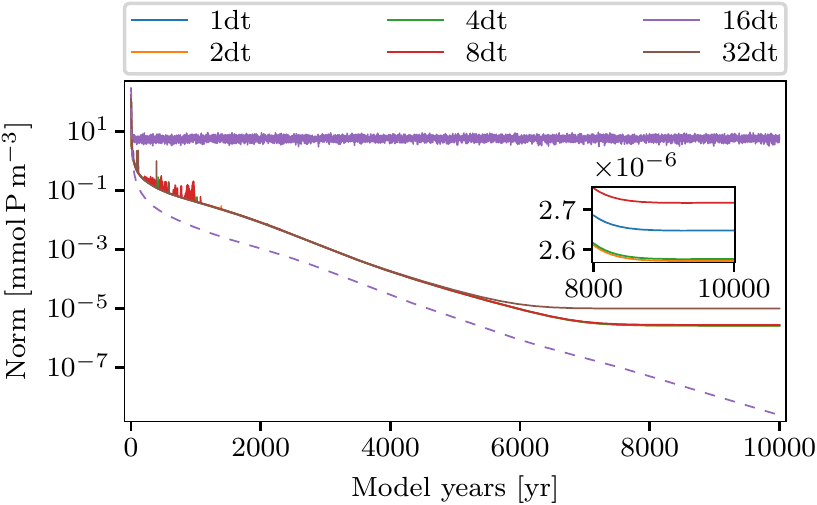}}
    \qquad
    \subfloat[NPZD-DOP model using parameter vector $\mathbf{u}_{\text{NPZD-DOP}}$
             \eqref{eqn:ParameterVector_NPZD-DOP}. \label{fig:Spinup-NPZD-DOP}]{\includegraphics{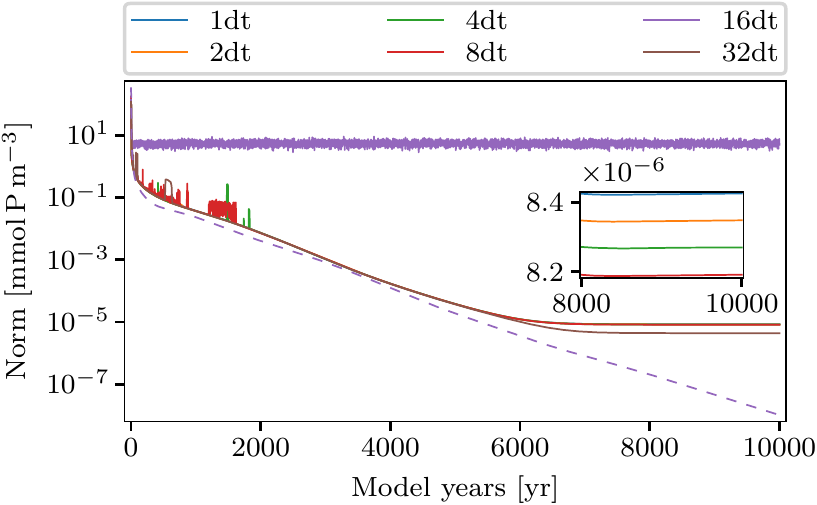}}
    \qquad
    \subfloat[MITgcm-PO4-DOP model using parameter vector $\mathbf{u}_{\text{ref}}$
             \eqref{eqn:ParameterVector_MITgcm-PO4-DOP}. \label{fig:Spinup-MITgcm-PO4-DOP}]{\includegraphics{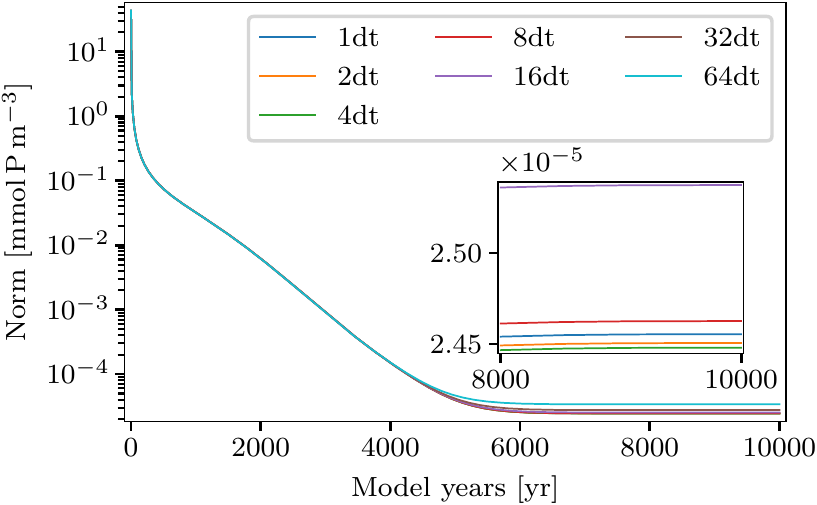}}
    \caption{Norm of the difference \eqref{eqn:StoppingCriterion} between two
             consecutive iterations in the spin-up for one parameter vector
             using different time steps (see Table \ref{table:Timesteps}).
             Additionally, the dashed line in Figs. \ref{fig:Spinup-NPZ-DOP}
             and \ref{fig:Spinup-NPZD-DOP} shows the norm of the differences
             \eqref{eqn:StoppingCriterion} for the spin-up where all tracers
             were initialized with a global mean concentration of
             \SI{0.542575}{\milli\mol\Phosphat\per\cubic\meter} (Fig. 
             \ref{fig:Spinup-NPZ-DOP}) and
             \SI{0.43408}{\milli\mol\Phosphat\per\cubic\meter} (Fig.
             \ref{fig:Spinup-NPZD-DOP}).}
    \label{fig:Spinup}
  \end{figure*}

  \begin{figure*}[ptb]
    \centering
    \subfloat[N model using parameter vector $\mathbf{u}_{\text{N}}$
             \eqref{eqn:ParameterVector_N}. \label{fig:Surface-N-N-64dt}]{\includegraphics{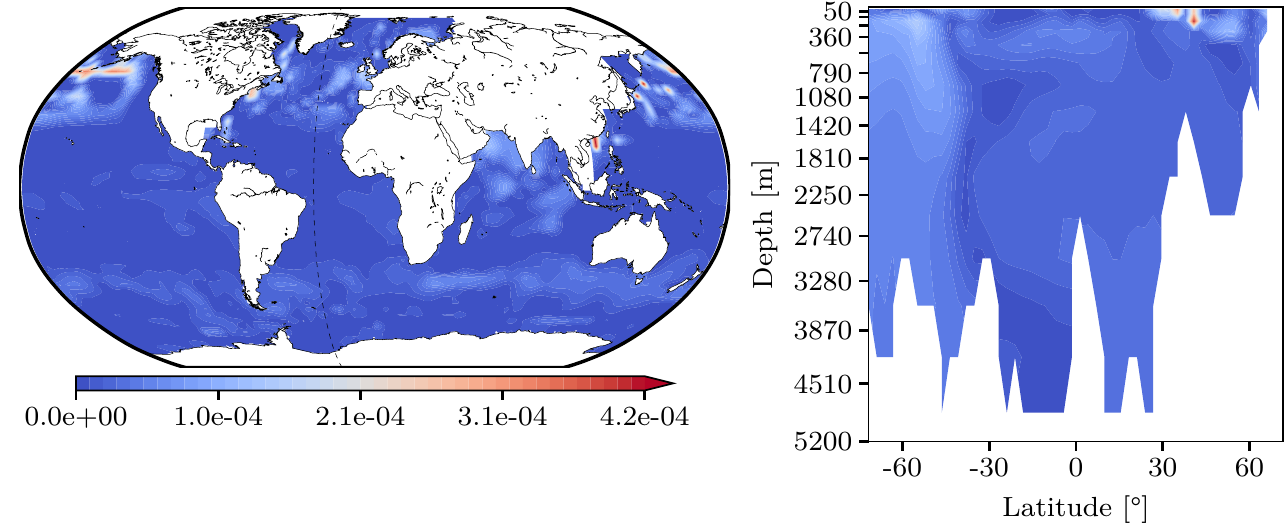}}
    \qquad
    \subfloat[N-DOP model using parameter vector $\mathbf{u}_{\text{N-DOP}}$
             \eqref{eqn:ParameterVector_N-DOP}. \label{fig:Surface-N-DOP-N-64dt}]{\includegraphics{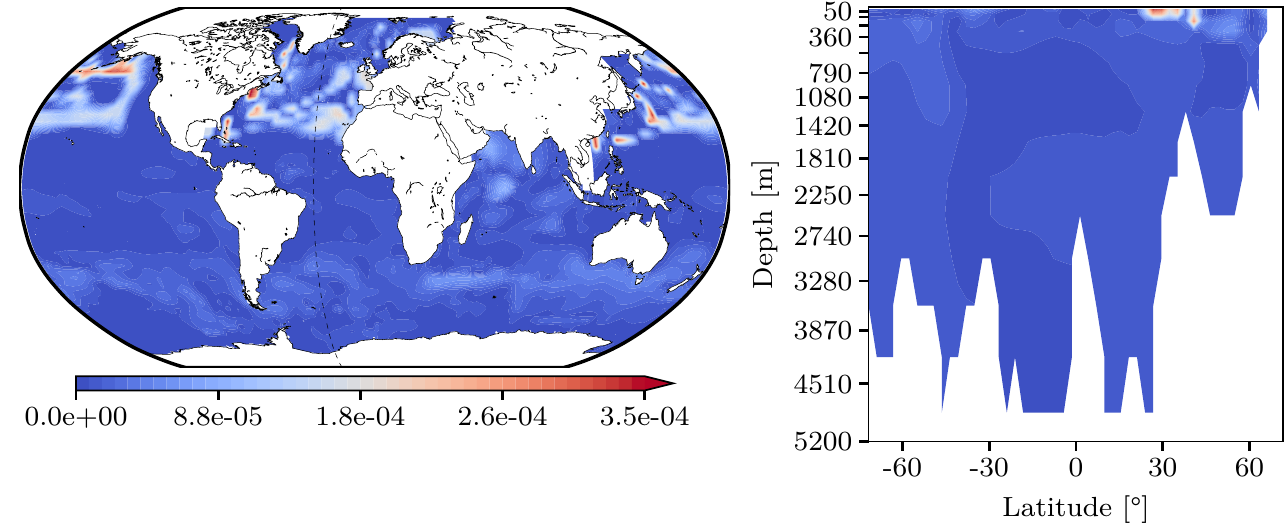}}
    \qquad
    \subfloat[MITgcm-PO4-DOP model using parameter vector $\mathbf{u}_{\text{ref}}$
             \eqref{eqn:ParameterVector_MITgcm-PO4-DOP}. \label{fig:Surface-PO4-MITgcm-PO4-DOP-64dt}]{\includegraphics{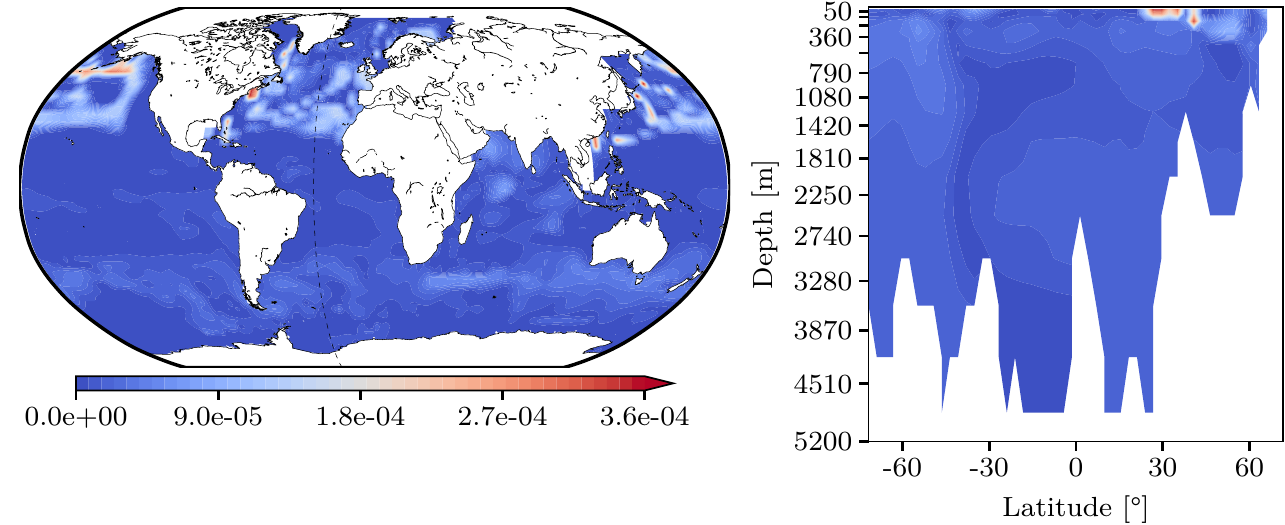}}
    \caption{Relative error \eqref{eqn:relativeErrorSurface} with
             $\ell = 10000$ and $t = 64$ of the phosphate concentrations
             at the surface layer (\SIrange{0}{50}{\metre}, left) and at a slice
             plane of the Atlantic at \ang{29.53125} W (located at the dashed
             line, right) at the first time instant of the model year (in
             January).}
    \label{fig:Surface-64dt}
  \end{figure*}

  \begin{figure*}[ptb]
    \centering
    \subfloat[NP-DOP model using parameter vector $\mathbf{u}_{\text{NP-DOP}}$
             \eqref{eqn:ParameterVector_NP-DOP}. \label{fig:Surface-NP-DOP-N-64dt}]{\includegraphics{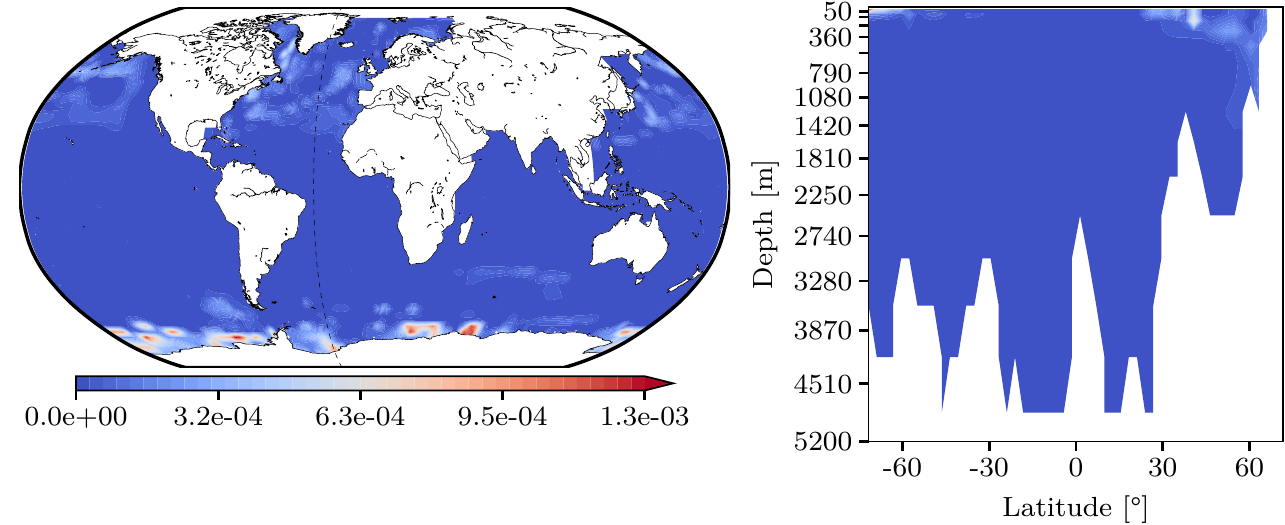}}
    \qquad
    \subfloat[NPZ-DOP model using parameter vector $\mathbf{u}_{\text{NPZ-DOP}}$
             \eqref{eqn:ParameterVector_NPZ-DOP}. \label{fig:Surface-NPZ-DOP-N-64dt}]{\includegraphics{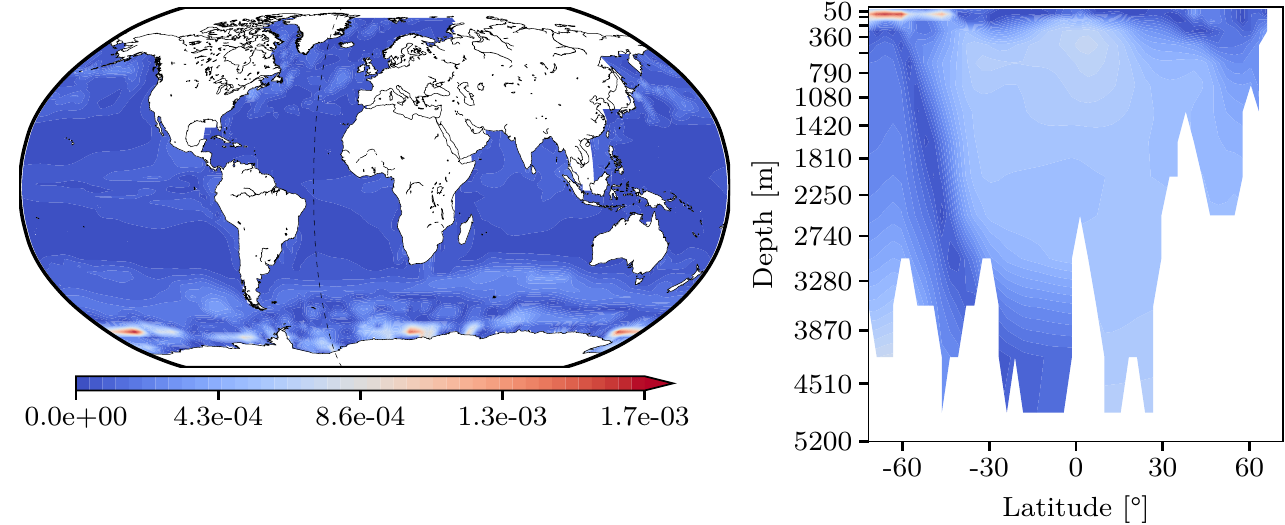}}
    \qquad
    \subfloat[NPZD-DOP model using parameter vector $\mathbf{u}_{\text{NPZD-DOP}}$
             \eqref{eqn:ParameterVector_NPZD-DOP}. \label{fig:Surface-NPZD-DOP-N-64dt}]{\includegraphics{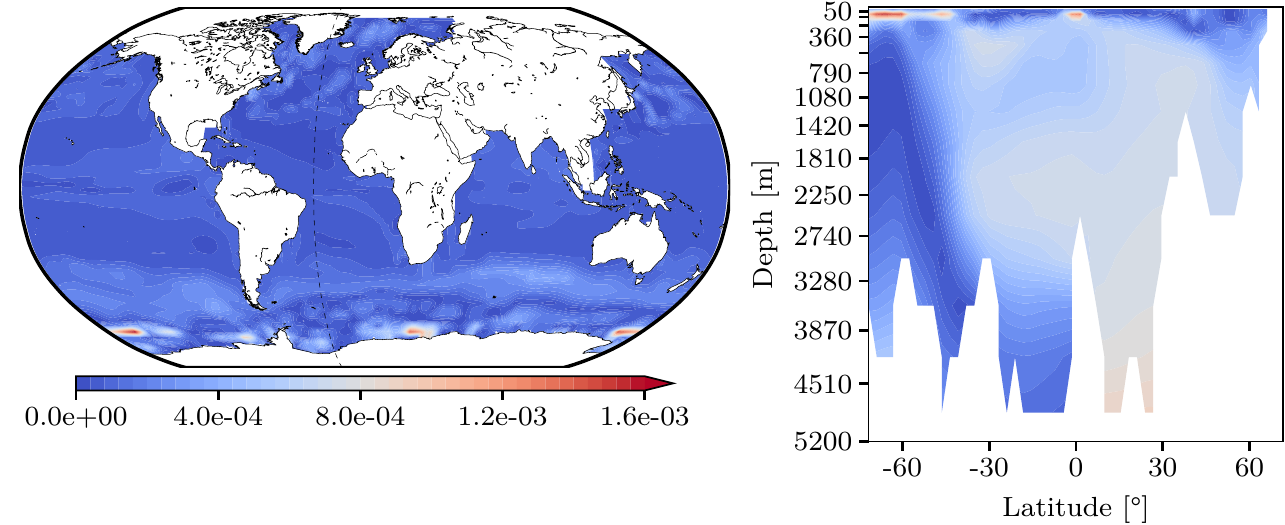}}
    \caption{As Fig. \ref{fig:Surface-64dt}, but for $t = 32$ and the other
             biogeochemical models.}
    \label{fig:Surface-32dt}
  \end{figure*}

  \begin{table*}[tb]
    \caption{Relative error \eqref{eqn:relativeError} with $\ell = 10000$ and
             $t \in \{2, 4, 8, 16, 32, 64\}$ for the whole model hierarchy.}
    \label{table:Norms-Modelhierarchy}
    \begin{tabular}{c c c c c c}
      \tophline
      Time step & N & N-DOP & NP-DOP & NPZ-DOP & NPZD-DOP \\
      \middlehline
      \SI{2}{\Timestep}  & 7.816e-05 & 7.994e-05 & 4.715e-04 & 5.537e-04 & 5.757e-04 \\
      \SI{4}{\Timestep}  & 2.239e-04 & 2.248e-04 & 1.277e-03 & 1.383e-03 & 1.398e-03 \\
      \SI{8}{\Timestep}  & 5.022e-04 & 5.106e-04 & 2.912e-03 & 2.997e-03 & 2.795e-03 \\
      \SI{16}{\Timestep} & 1.084e-03 & 1.214e-03 & 6.212e-03 & 1.942e-02 & 1.722e-02 \\
      \SI{32}{\Timestep} & 2.259e-03 & 2.775e-03 & 1.355e-02 & 5.513e-02 & 6.146e-02 \\
      \SI{64}{\Timestep} & 4.035e-03 & 4.923e-03 &     -     &     -     &     -     \\
      \bottomhline
    \end{tabular}
  \end{table*}

  The spin-ups calculated with different time steps reflected the same steady
  annual cycle approximately. We first analyzed the influence of the time step
  on the accuracy of the steady annual cycle using one parameter vector for each
  biogeochemical model. We used the parameter vectors
  \begin{figure*}[ptb]
    \centering
    \subfloat[N model using parameter vector $\mathbf{u}_{\text{N}}$
             \eqref{eqn:ParameterVector_N}. \label{fig:Norm-2-N}]{\includegraphics{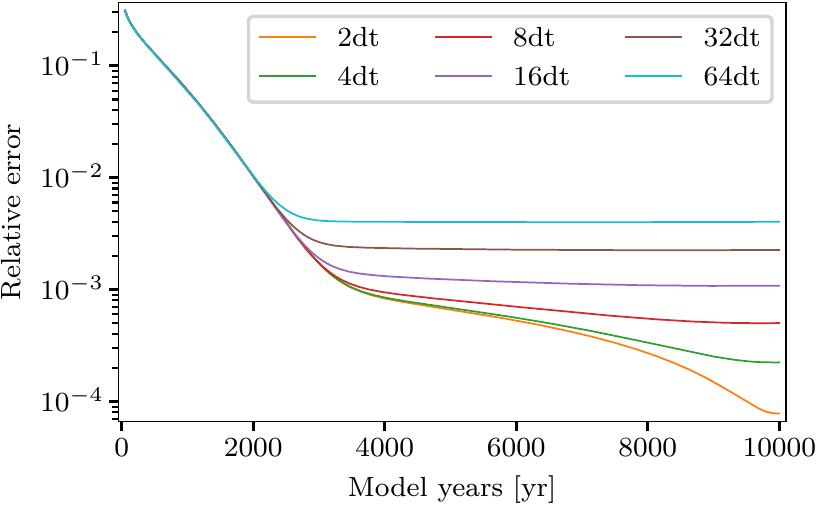}}
    \qquad
    \subfloat[N-DOP model using parameter vector $\mathbf{u}_{\text{N-DOP}}$
             \eqref{eqn:ParameterVector_N-DOP}. \label{fig:Norm-2-N-DOP}]{\includegraphics{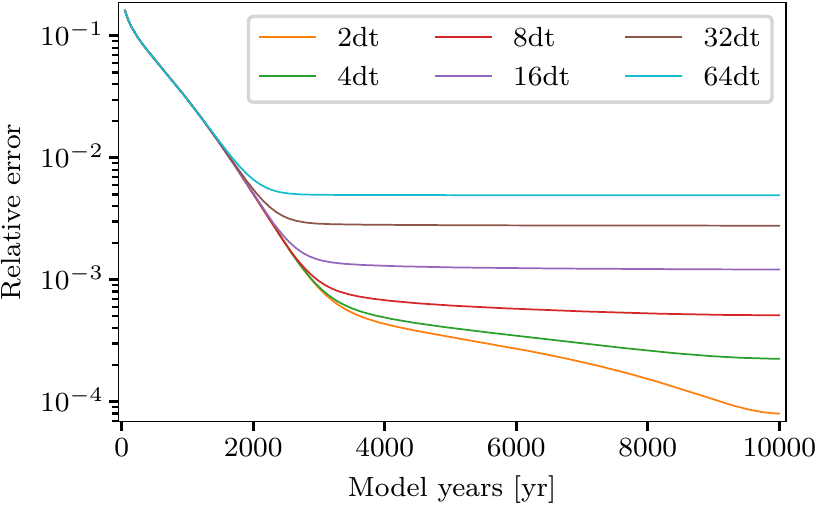}}
    \qquad
    \subfloat[NP-DOP model using parameter vector $\mathbf{u}_{\text{NP-DOP}}$
             \eqref{eqn:ParameterVector_NP-DOP}. \label{fig:Norm-2-NP-DOP}]{\includegraphics{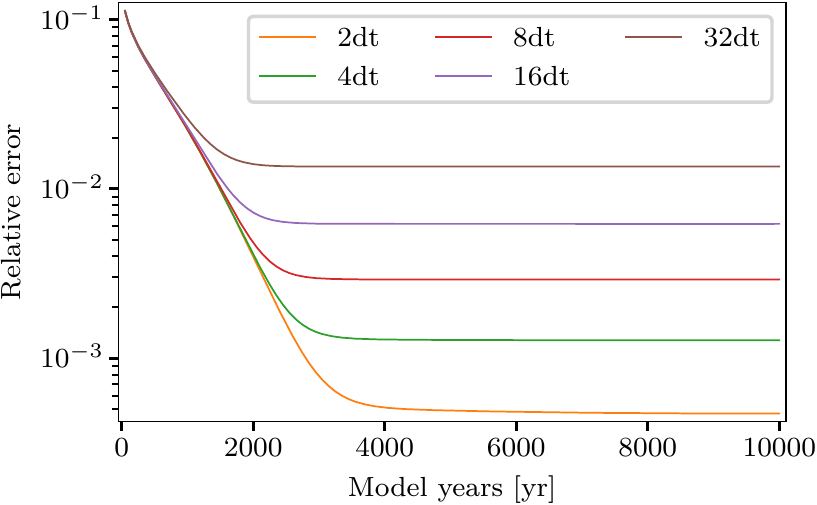}}
    \qquad
    \subfloat[NPZ-DOP model using parameter vector $\mathbf{u}_{\text{NPZ-DOP}}$
             \eqref{eqn:ParameterVector_NPZ-DOP}. \label{fig:Norm-2-NPZ-DOP}]{\includegraphics{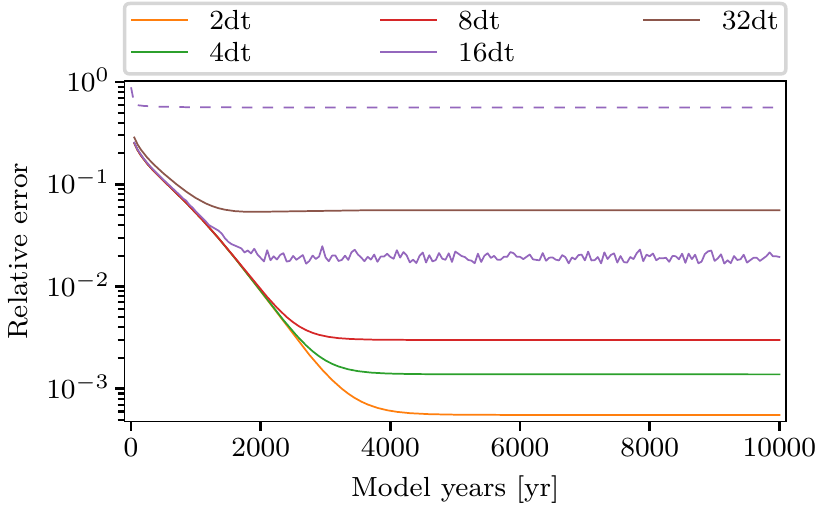}}
    \qquad
    \subfloat[NPZD-DOP model using parameter vector $\mathbf{u}_{\text{NPZD-DOP}}$
             \eqref{eqn:ParameterVector_NPZD-DOP}. \label{fig:Norm-2-NPZD-DOP}]{\includegraphics{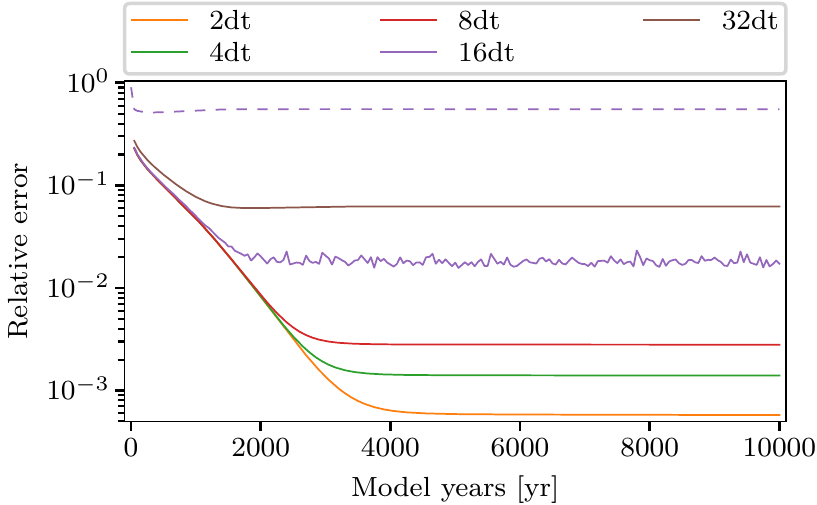}}
    \qquad
    \subfloat[MITgcm-PO4-DOP model using parameter vector $\mathbf{u}_{\text{ref}}$
             \eqref{eqn:ParameterVector_MITgcm-PO4-DOP}. \label{fig:Norm-2-MITgcm-PO4-DOP}]{\includegraphics{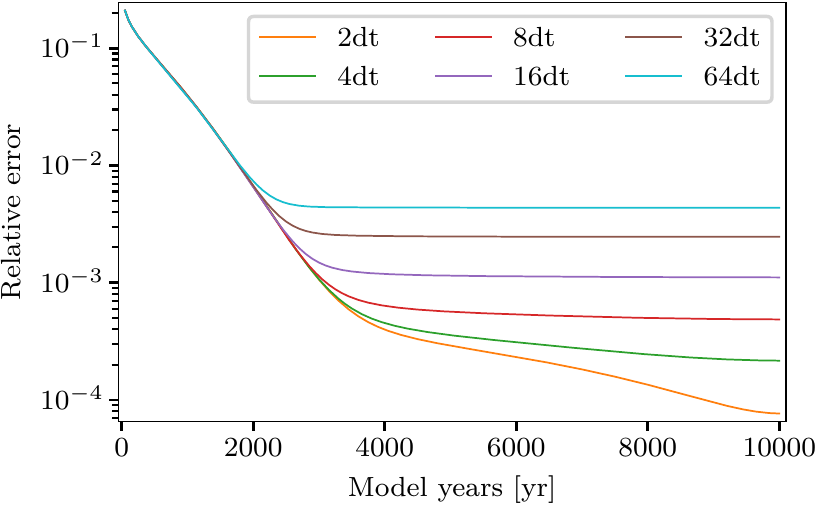}}
    \caption{Relative error \eqref{eqn:relativeError} in the Euclidean norm of
             the spin-up using different time steps for $\ell \in \{50, 100,
             \ldots, 10000\}$, $t \in \{2, 4, 8, 16, 32, 64\}$ and one parameter
             vector respectively. The dashed line represents the relative error
             calculated with a global mean concentration of
             \SI{0.542575}{\milli\mol\Phosphat\per\cubic\meter} (Fig.
             \ref{fig:Norm-2-NPZ-DOP}) and
             \SI{0.43408}{\milli\mol\Phosphat\per\cubic\meter} (Fig.
             \ref{fig:Norm-2-NPZD-DOP}) for each tracer.}
    \label{fig:Norm-2}
  \end{figure*}
  \begin{align}
    \label{eqn:ParameterVector_N}
    \mathbf{u}_{\text{N}} &= \left( 0.02, 2.0, 0.5, 30.0, 0.858 \right) \\
    \label{eqn:ParameterVector_N-DOP}
    \mathbf{u}_{\text{N-DOP}} &= \left( 0.02, 2.0, 0.5, 30.0, 0.67, 0.5, 0.858
                                 \right) \\
    \label{eqn:ParameterVector_NP-DOP}
    \mathbf{u}_{\text{NP-DOP}} &= \left( 0.02, 0.48, 2.0, 2.0, 0.5, 0.088, 30.0,
                                         0.67, 0.04, 4.0, 0.01, 0.5, 0.858
                                  \right) \\
    \label{eqn:ParameterVector_NPZ-DOP}
    \mathbf{u}_{\text{NPZ-DOP}} &= \left( 0.02, 0.48, 2.0, 2.0, 0.5, 0.088,
                                          30.0, 0.75, 0.67, 0.04, 0.03, 3.2,
                                          0.01, 0.01, 0.5, 0.858
                                   \right) \\
    \label{eqn:ParameterVector_NPZD-DOP}
    \mathbf{u}_{\text{NPZD-DOP}} &= \left( 0.02, 0.48, 2.0, 2.0, 0.5, 0.088,
                                          30.0, 0.75, 0.67, 0.04, 0.03, 3.2,
                                          0.01, 0.01, 0.05, 0.5, 0.058, 0.0
                                   \right) \\
    \label{eqn:ParameterVector_MITgcm-PO4-DOP}
     \mathbf{u}_{\text{ref}} &= \left( 0.5, 2.0, 0.67, 0.5, 30.0, 0.02,
                                      0.858 \right)
  \end{align}
  (see Tables \ref{table:Parameter-Modelhierarchy} and
  \ref{table:Parameter-MITgcm-PO4-DOP}). Figure \ref{fig:Spinup} demonstrates a
  similar convergence behavior for all different time steps whereas the spin-up
  reached nearly the same accuracy for the norm of differences 
  \eqref{eqn:StoppingCriterion}. For the NP-DOP, NPZ-DOP and NPZD-DOP model, the
  spin-up calculation using time step \SI{64}{\Timestep} was aborted due to
  a technical error as consequence of a too large step size of the Euler method.
  Moreover, the norm of differences oscillated and the spin-up did not converge
  when using time step \SI{16}{\Timestep} for the NPZ-DOP and NPZD-DOP model.
  For all biogeochemical models, the magnitude of the error
  \eqref{eqn:relativeErrorSurface} increased when larger time steps were used
  (for the models N, N-DOP and MITgcm-PO4-DOP from $10^{-6}$ (\SI{2}{\Timestep})
  to $10^{-3}$ (\SI{64}{\Timestep}) and the same also for the models NP-DOP,
  NPZ-DOP and NPZD-DOP from $10^{-5}$ (\SI{2}{\Timestep}) to $10^{-3}$
  (\SI{32}{\Timestep})) (Figs. \ref{fig:Surface-64dt} and \ref{fig:Surface-32dt}).
  With each model, the largest errors occurred almost in the same regions for
  the different time steps. Likewise, this was observed with the other tracers.
  More specifically, the accuracy of the calculated steady annual cycles
  decreased not only on the surface layer but also in the entire ocean when
  larger time steps were used as shown in Fig. \ref{fig:Norm-2} and Table
  \ref{table:Norms-Modelhierarchy}. Namely, the accuracy developed
  equally for all time steps for the first model years but the error reduction
  stagnated earlier with increasing time steps during the spin-up. Although the
  norm of the differences \eqref{eqn:StoppingCriterion} oscillated and did not
  converge using time step \SI{16}{\Timestep} for the NPZ-DOP and NPZD-DOP
  model (Figs. \ref{fig:Spinup-NPZ-DOP} and \ref{fig:Spinup-NPZD-DOP}), the
  relative error was between those calculated with time steps \SI{8}{\Timestep}
  and \SI{32}{\Timestep}. The choice of the norm (i.e., volume-weighted
  norm $\left\| \cdot \right\|_{2, V}$ or comparing the whole trajectory
  $\left\| \cdot \right\|_{2,V,T}$) did not lead to any qualitative difference
  in the calculation of the relative error \eqref{eqn:relativeError} (Table
  \ref{table:Norms-MITgcm-PO4-DOP}). The cost function values in Table 
  \ref{table:Costfunction-Modelhierarchy} indicate that the steady
  annual cycle when calculated with larger time steps reflected the data of the
  world ocean database in a similar manner as the steady annual cycle calculated
  with time step \SI{1}{\Timestep}. However, the cost function values were quite
  large for all models and time steps because the parameter vectors have not
  been optimized for the steady state to reflect the data optimally.

  \begin{table*}[tb]
    \caption{The use of different norms for the relative error
             \eqref{eqn:relativeError} with $\ell = 10000$ and $t \in \{2, 4,
             8, 16, 32, 64\}$ for the MITgcm-PO4-DOP model and parameter
             vector $\mathbf{u}_{\text{ref}}$
             \eqref{eqn:ParameterVector_MITgcm-PO4-DOP}.}
    \label{table:Norms-MITgcm-PO4-DOP}
    \begin{tabular}{c c c c c c}
      \tophline
      Time step & $\left\| \cdot \right\|_2$ & $\left\| \cdot \right\|_{2, V}$ & $\left\| \cdot \right\|_{2,V,T}$ \\
      \middlehline
      \SI{2}{\Timestep}  & 7.680e-05 & 4.007e-05 & 7.008e-05 \\
      \SI{4}{\Timestep}  & 2.168e-04 & 1.158e-04 & 2.845e-04 \\
      \SI{8}{\Timestep}  & 4.872e-04 & 2.594e-04 & 8.741e-04 \\
      \SI{16}{\Timestep} & 1.111e-03 & 5.587e-04 & 2.377e-03 \\
      \SI{32}{\Timestep} & 2.462e-03 & 1.166e-03 & 6.032e-03 \\
      \SI{64}{\Timestep} & 4.360e-03 & 2.061e-03 & 1.497e-02 \\
      \bottomhline
    \end{tabular}
  \end{table*}

  \begin{table*}[t]
    \caption{Cost function values of type ordinary least squares
             \eqref{eqn:Costfunction} for the whole model hierarchy. Shown are
             the cost functions values $J_{\text{OLS}} \left(\vec{y}^{10000, t}
             \right)$ of the spin-up using the exemplary parameter vectors (cf.
             Tables \ref{table:Parameter-Modelhierarchy} and
             \ref{table:Parameter-MITgcm-PO4-DOP}) for $t \in \{1, 2, 4, 8, 16,
             32, 64\}$, respectively. For the calculation of the cost function,
             only the N tracer was used because hardly any measurement data are
             available for the other tracers.}
    \label{table:Costfunction-Modelhierarchy}
    \begin{tabular}{c c c c c c c}
      \tophline
      Time step & N & N-DOP & NP-DOP & NPZ-DOP & NPZD-DOP & MITgcm-PO4-DOP \\
      \middlehline
      \SI{1}{\Timestep}  & 1.093e+06 & 9.834e+05 & 1.261e+06 & 1.274e+06 & 1.232e+06 & 8.375e+05 \\
      \SI{2}{\Timestep}  & 1.110e+06 & 9.930e+05 & 1.268e+06 & 1.280e+06 & 1.235e+06 & 8.473e+05 \\
      \SI{4}{\Timestep}  & 2.428e+06 & 1.471e+06 & 1.558e+06 & 1.805e+06 & 1.551e+06 & 1.297e+06 \\
      \SI{8}{\Timestep}  & 2.927e+06 & 1.626e+06 & 1.665e+06 & 1.999e+06 & 1.673e+06 & 1.456e+06 \\
      \SI{16}{\Timestep} & 1.657e+06 & 1.315e+06 & 1.507e+06 & 1.590e+06 & 1.499e+06 & 1.223e+06 \\
      \SI{32}{\Timestep} & 1.686e+06 & 1.402e+06 & 1.609e+06 & 1.626e+06 & 1.480e+06 & 1.292e+06 \\
      \SI{64}{\Timestep} & 1.713e+06 & 1.490e+06 &     -     &     -     &     -     & 1.366e+06 \\
  \bottomhline
  \end{tabular}
  \end{table*}

  The reason for the divergence of the solution using time step
  \SI{64}{\Timestep} was the too large step size for the Euler method. The
  biogeochemical model, for example, used more nutrients than available in a box
  of the spatial discretization for one time step and converted these into
  particle like plankton due to the large time step. Thus, the concentration in
  this box became negative, i.e., the model ran empty. As a consequence, the
  model produced mass because the biogeochemical model set the negative
  concentration value to zero but did not subtract the absolute value anywhere
  else. On the contrary, the biogeochemical model did not generate negative
  concentrations because the equations of this model are quasi-positive
  \citep[cf.][]{Pie10}, and the transport matrix method also does not generate
  negative concentrations. Indeed, the negative concentrations occurred due to
  the too large step size of the Euler method and, subsequently, caused the
  increase of absolute value of the negative concentrations and, as a result, of
  the positive concentrations during the spin-up. Therefore, NaN (not a number)
  arose in the computation of the quadratic loss term for phytoplankton in the
  zooplankton grazing \eqref{eqn:Zooplankton} due to the large phytoplankton
  concentration which became infinite in the quadratic term of the numerator as
  well as that of the denominator.

  The oscillation of the difference norm \eqref{eqn:StoppingCriterion} resulted
  from an inappropriate step size of the Euler method (Figs.
  \ref{fig:Spinup-NPZ-DOP} and \ref{fig:Spinup-NPZD-DOP}). For the spin-up
  calculation with \SI{16}{\Timestep}, the oscillation occurred for each
  tracer. The oscillation started on the ocean surface and spread to deeper
  layers during the spin-up so that the oscillation appeared in the lowest
  layer between \SIrange{4510}{5200}{\meter} only after \num{2000} model years.
  The magnitude of the norm of differences, and consequently of the
  oscillation, decreased on deeper layers because the annual concentration
  change on deeper layers is marginal. Apart from that, the occurrence of
  oscillations depended on the initial tracer concentrations (Figs.
  \ref{fig:Spinup-NPZ-DOP} and \ref{fig:Spinup-NPZD-DOP}, dashed line). For the
  changed initial concentration, the spin-up, however, ended with an invalid
  steady annual cycle (Figs. \ref{fig:Norm-2-NPZ-DOP} and
  \ref{fig:Norm-2-NPZD-DOP}, dashed line) because, first, all tracers had nearly
  constant concentrations, second, all the mass was concentrated in the \unit{N}
  tracer, and third, the other tracers (\unit{P}, \unit{Z} and \unit{DOP}) had
  negative concentrations exclusively. This was a consequence of a too large
  time step for the Euler method.
  
  In short, all steady annual cycles calculated with larger time steps
  practically approximated the reference solution for the whole model
  hierarchy. The accordance between the reference solution and the steady
  annual cycle computed with larger time steps decreased slowly with larger time
  steps, on the one hand, and, on the other hand, slightly with the complexity
  of the biogeochemical model as detailed in Table
  \ref{table:Norms-Modelhierarchy}.

\subsection{Parameter samples}
\label{sec:Results-ParameterSamples}

  In this section, we analyzed the behavior for the different time steps
  utilizing the parameter vectors of the Latin hypercube sample for all
  biogeochemical models. In the previous Sect.
  \ref{sec:Results-LargerTimeSteps}, we discussed this behavior exclusively
  for one (reference) parameter vector for each model.

  \begin{figure*}[p]
    \centering
    \subfloat[N model\label{fig:Lhs-SpinupNorm-RelError-N}]{\includegraphics{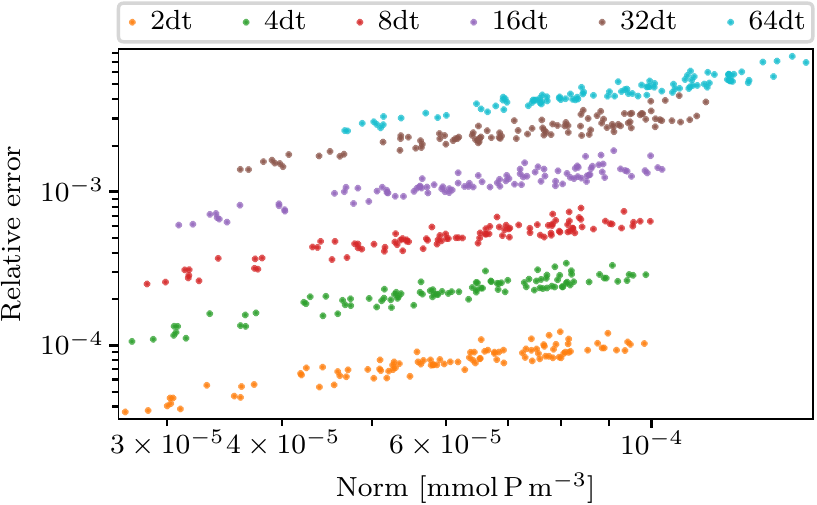}}
    \qquad
    \subfloat[N-DOP model\label{fig:Lhs-SpinupNorm-RelError-N-DOP}]{\includegraphics{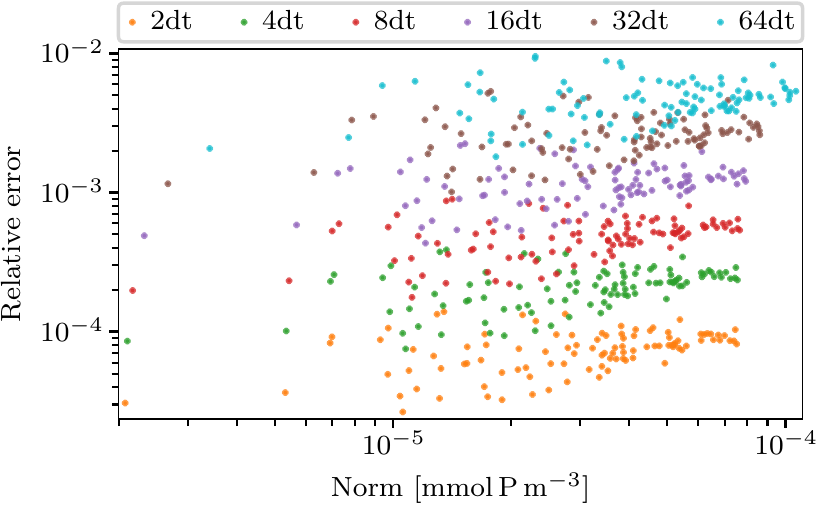}}
    \qquad
    \subfloat[NP-DOP model\label{fig:Lhs-SpinupNorm-RelError-NP-DOP}]{\includegraphics{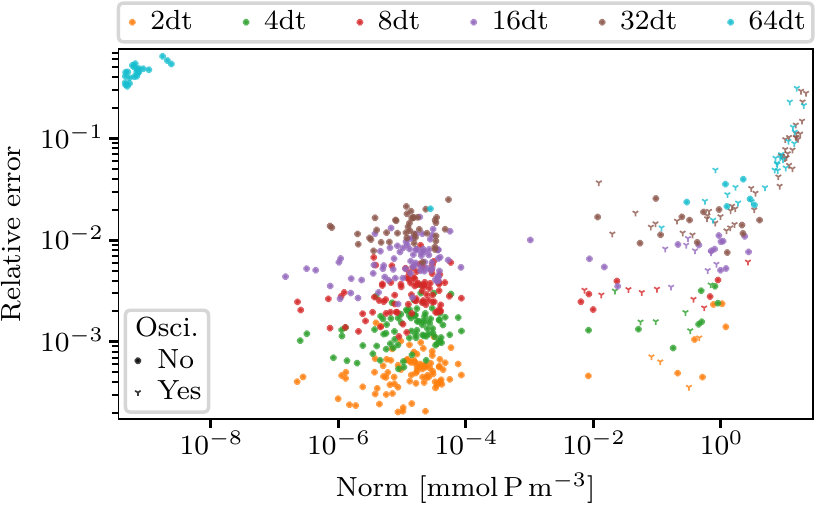}}
    \qquad
    \subfloat[NPZ-DOP model\label{fig:Lhs-SpinupNorm-RelError-NPZ-DOP}]{\includegraphics{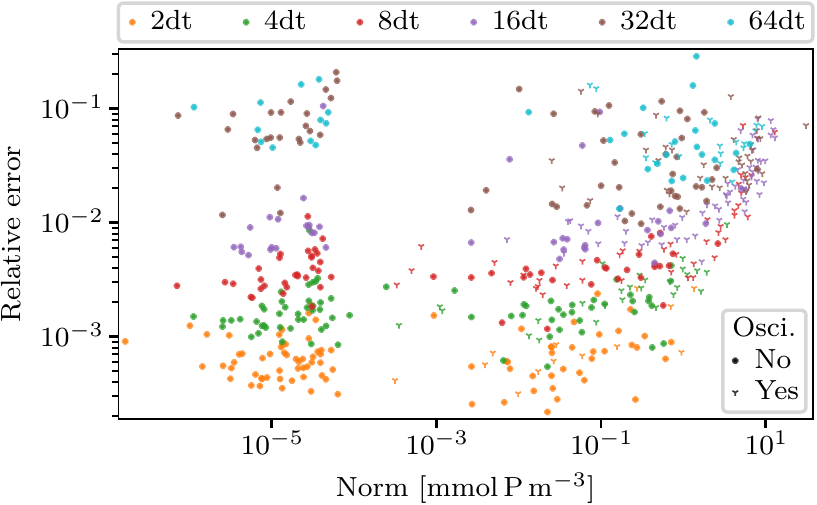}}
    \qquad
    \subfloat[NPZD-DOP model\label{fig:Lhs-SpinupNorm-RelError-NPZD-DOP}]{\includegraphics{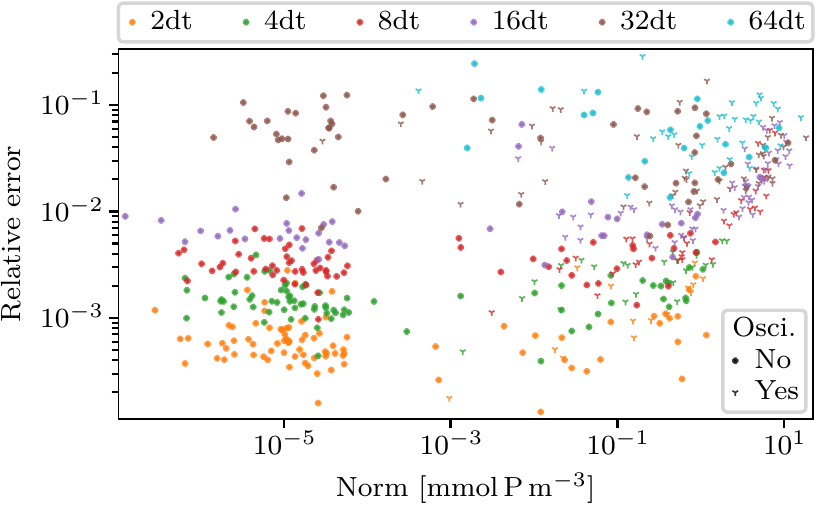}}
    \qquad
    \subfloat[MITgcm-PO4-DOP model\label{fig:Lhs-SpinupNorm-RelError-MITgcm-PO4-DOP}]{\includegraphics{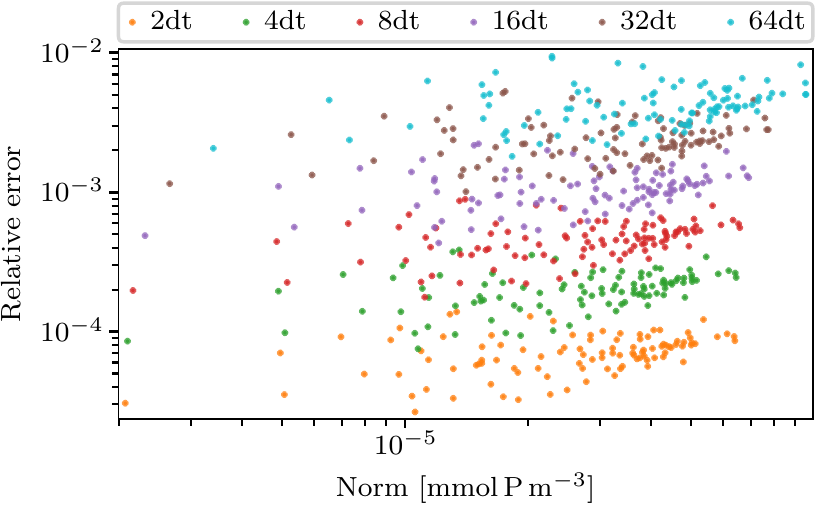}}
    \caption{Visualization of the norm of the differences
             \eqref{eqn:StoppingCriterion} and the relative error
             \eqref{eqn:relativeError} in the Euclidean norm for $\ell = 10000$
             using different time steps for all parameter vectors of the Latin 
             hypercube sample.}
    \label{fig:Lhs-SpinupNorm-RelError}
  \end{figure*}

  \begin{table*}[tb]
    \caption{Count of parameter sets of the Latin hypercube sample without
             spin-up convergence after \num{10000} model years for different time
             steps.}
    \label{table:LHS-noSpinup}
    \begin{tabular}{c c c c c c c}
      \tophline
      Time step & N & N-DOP & NP-DOP & NPZ-DOP & NPZD-DOP & MITgcm-PO4-DOP \\
      \middlehline
       1 \unit{dt} &  0 &  0 &  0 &  0 &  0 &  0 \\
       2 \unit{dt} &  0 &  0 &  0 &  0 &  0 &  0 \\
       4 \unit{dt} &  0 &  0 &  0 &  0 &  0 &  0 \\
       8 \unit{dt} &  0 &  0 &  0 &  0 &  0 &  0 \\
      16 \unit{dt} &  0 &  0 &  0 &  0 &  0 &  0 \\
      32 \unit{dt} &  0 &  0 &  0 &  5 &  3 &  0 \\
      64 \unit{dt} &  0 &  0 & 45 & 47 & 46 &  0 \\
      \bottomhline
    \end{tabular}
  \end{table*}

  \begin{table*}[tb]
    \caption{Count of parameter sets of the Latin hypercube sample with
             oscillations of the norm of differences
             \eqref{eqn:StoppingCriterion} during the spin-up.}
    \label{table:LHS-Oscillations}
    \begin{tabular}{c c c c c c c}
      \tophline
      Time step & N & N-DOP & NP-DOP & NPZ-DOP & NPZD-DOP & MITgcm-PO4-DOP \\
      \middlehline
       1 \unit{dt} &  0 &   0 &   5 &  12 &  14 &   0 \\
       2 \unit{dt} &  0 &   0 &   4 &  13 &  14 &   0 \\
       4 \unit{dt} &  0 &   0 &   6 &  27 &  23 &   0 \\
       8 \unit{dt} &  0 &   0 &   8 &  43 &  34 &   0 \\
      16 \unit{dt} &  0 &   0 &   8 &  60 &  62 &   0 \\
      32 \unit{dt} &  0 &   0 &  45 &  36 &  46 &   0 \\
      64 \unit{dt} &  0 &   0 &  27 &  19 &  35 &   0 \\
      \bottomhline
    \end{tabular}
  \end{table*}

  The accuracy of the approximation of the steady annual cycle decreased with
  larger time steps. For the three biogeochemical models N, N-DOP and
  MITgcm-PO4-DOP, the spin-up calculations almost always converged with a norm
  of differences \eqref{eqn:StoppingCriterion} less than $10^{-4}$ whereby the
  difference increased only slightly for larger time steps (Figs.
  \ref{fig:Lhs-SpinupNorm-RelError-N}, \ref{fig:Lhs-SpinupNorm-RelError-N-DOP}
  and \ref{fig:Lhs-SpinupNorm-RelError-MITgcm-PO4-DOP}). In contrast, the
  spin-ups for the other three models NP-DOP, NPZ-DOP and NPZD-DOP ended up with
  a much larger norm of differences between two successive iterations  
  \eqref{eqn:StoppingCriterion} for numerous parameter vectors (Figs.
  \ref{fig:Lhs-SpinupNorm-RelError-NP-DOP},
  \ref{fig:Lhs-SpinupNorm-RelError-NPZ-DOP} and
  \ref{fig:Lhs-SpinupNorm-RelError-NPZD-DOP}). As a result of a too large time
  step, the spin-up calculation diverged for almost \num{50}\% of the parameter
  vectors when using time step \SI{64}{\Timestep} (Table
  \ref{table:LHS-noSpinup}). For one half of these parameter vectors, the
  spin-up did not converge for all three models. For the parameter vectors of
  the other half, either the spin-up diverged exclusively for the NP-DOP model
  or did not converge for the NPZ-DOP and NPZD-DOP model. Furthermore, the 
  spin-up using time step \SI{64}{\Timestep} was always divergent if the
  spin-up calculation diverged using time step \SI{32}{\Timestep} for the
  same model and parameter vector. In addition, oscillations of the norm of
  differences \eqref{eqn:StoppingCriterion} occurred frequently when time steps
  larger than \SI{4}{\Timestep} for these three models were used as detailed in
  Table \ref{table:LHS-Oscillations}. There were parameter vectors for which
  oscillations occurred for exactly one time step as well as parameter vectors
  for which oscillations appeared for several (up to even all) time steps. The
  values of the individual model parameter of the parameter vectors covered the
  entire range in each case so that the oscillations could not be traced back to
  the value of an individual model parameter. Nonetheless, Fig.
  \ref{fig:Lhs-SpinupNorm-RelError} shows the approximately identical magnitude
  of the relative error \eqref{eqn:relativeError} for each time step, regardless
  of both the norm of differences \eqref{eqn:StoppingCriterion} and the
  occurrence of oscillations. Analogously to the results for the reference
  parameter vector (see Sect. \ref{sec:Results-LargerTimeSteps}), the relative
  error increased, on the one hand, with larger time steps and, on the other
  hand, with the complexity of the biogeochemical model. Moreover, the steady
  annual cycles matched the data of the world ocean database in a similar way
  for all time steps. As for the reference parameters (see Table
  \ref{table:Costfunction-Modelhierarchy}), the cost function values
  \eqref{eqn:Costfunction} calculated with time step \SI{1}{\Timestep} nearly
  corresponded to those calculated with \SI{2}{\Timestep} whereas the values
  calculated with larger time steps were slightly higher.

  \begin{figure*}[p]
    \centering
    \subfloat[N model \label{fig:Lhs-RequiredModelYears-RelError-N}]{\includegraphics{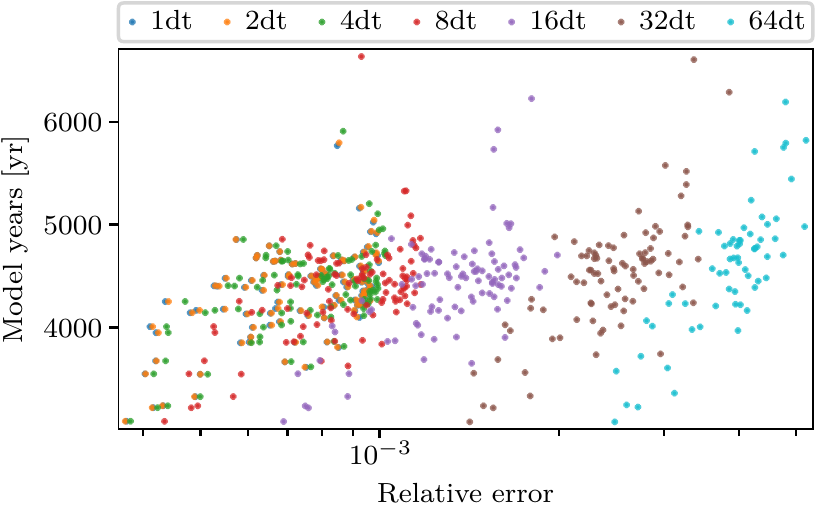}}
    \qquad
    \subfloat[N-DOP model \label{fig:Lhs-RequiredModelYears-RelError-N-DOP}]{\includegraphics{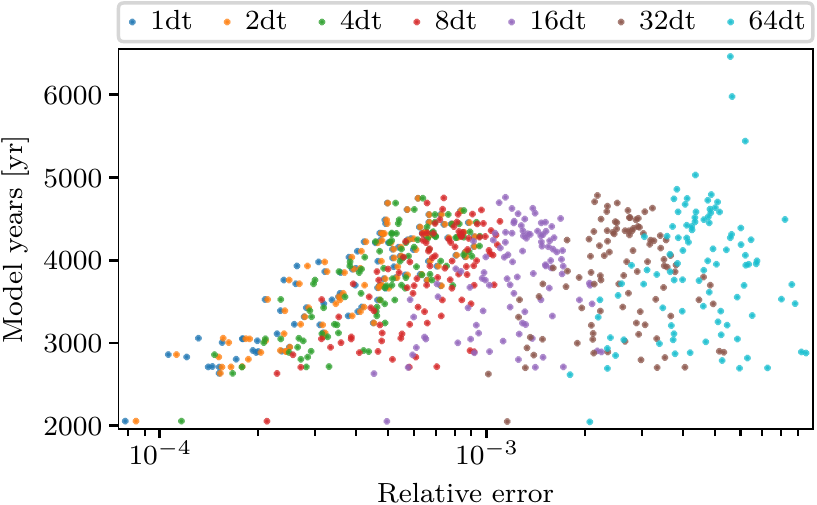}}
    \qquad
    \subfloat[NP-DOP model \label{fig:Lhs-RequiredModelYears-RelError-NP-DOP}]{\includegraphics{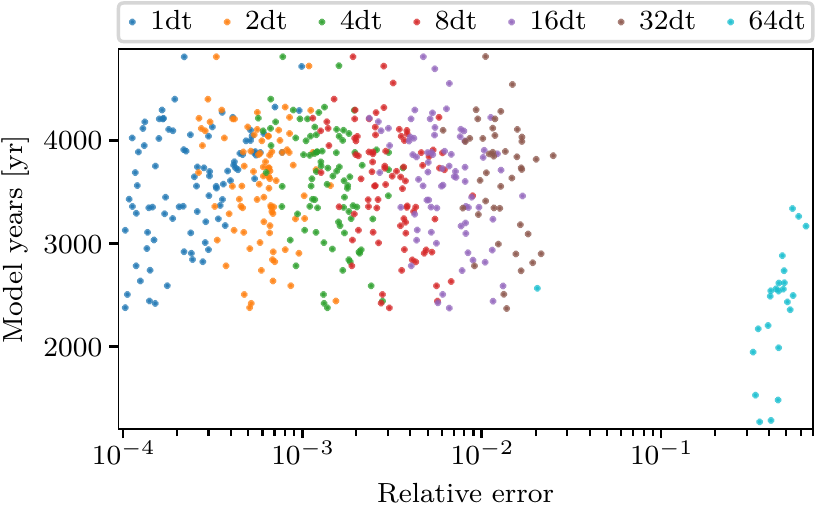}}
    \qquad
    \subfloat[NPZ-DOP model \label{fig:Lhs-RequiredModelYears-RelError-NPZ-DOP}]{\includegraphics{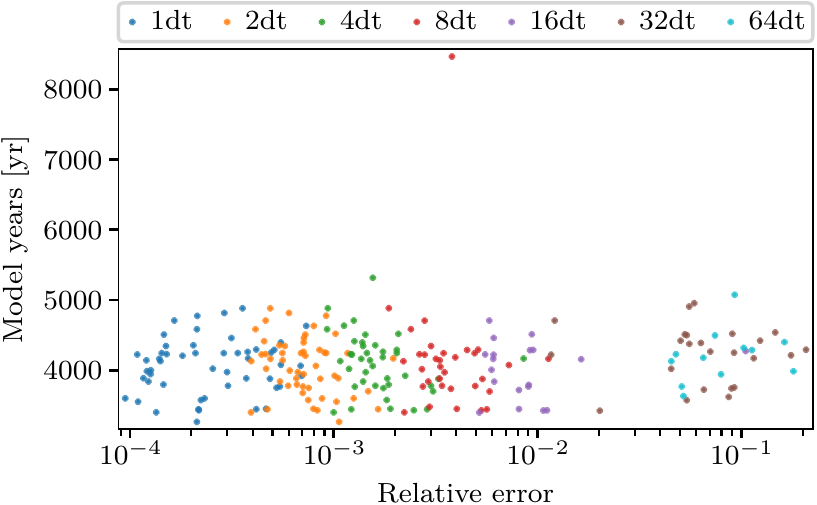}}
    \qquad
    \subfloat[NDPZ-DOP model \label{fig:Lhs-RequiredModelYears-RelError-NPZD-DOP}]{\includegraphics{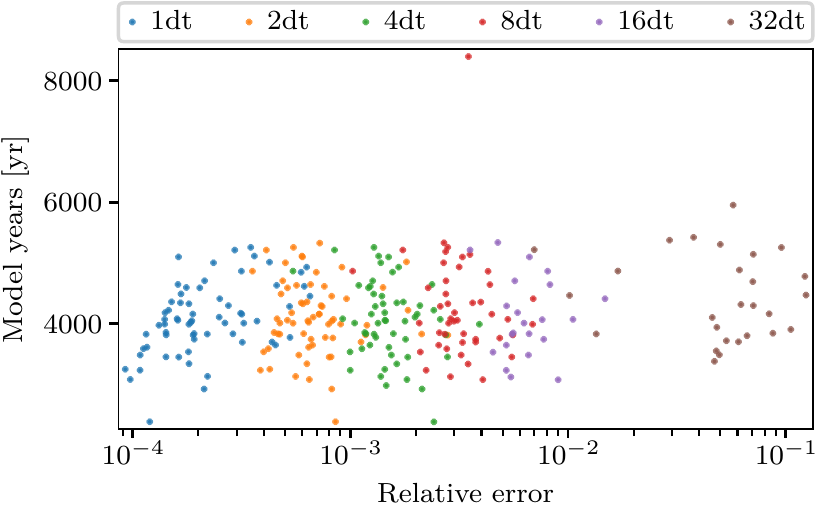}}
    \qquad
    \subfloat[MITgcm-PO4-DOP model \label{fig:Lhs-RequiredModelYears-RelError-MITgcm-PO4-DOP}]{\includegraphics{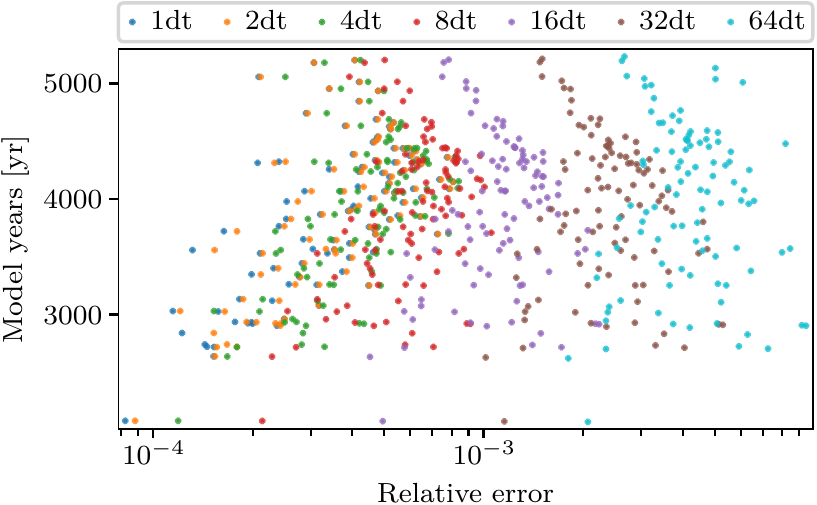}}
    \caption{Required model years for reaching a tolerance $10^{-4}$ in 
             \eqref{eqn:StoppingCriterion} in the spin-up and relative error
             \eqref{eqn:relativeError} of the corresponding model year using
             different time steps for all parameter vectors of the Latin
             hypercube sample.}
    \label{fig:Lhs-RequiredModelYears-RelError}
  \end{figure*}

  \begin{table*}[t]
    \caption{Count of parameter sets of the Latin hypercube sample, which reached
             a spin-up tolerance of less than $10^{-4}$ \unit{mmol\, P\, m^{-3}}
             after $10000$ model years, for the whole model hierarchy and
             different time steps.}
    \label{table:LHS-Spinup}
    \begin{tabular}{c c c c c c c}
      \tophline
      Time step & N & N-DOP & NP-DOP & NPZ-DOP & NPZD-DOP & MITgcm-PO4-DOP \\
      \middlehline
       1 \unit{dt} & 100 & 100 &  90 &  51 &  64 & 100 \\
       2 \unit{dt} & 100 & 100 &  89 &  52 &  64 & 100 \\
       4 \unit{dt} & 100 & 100 &  86 &  40 &  54 & 100 \\
       8 \unit{dt} & 100 & 100 &  86 &  32 &  45 & 100 \\
      16 \unit{dt} &  98 & 100 &  77 &  19 &  22 & 100 \\
      32 \unit{dt} &  89 & 100 &  42 &  23 &  26 & 100 \\
      64 \unit{dt} &  61 &  96 &  24 &  12 &   0 & 100 \\
      \bottomhline
    \end{tabular}
  \end{table*}

  The use of larger time steps significantly reduced the computational effort
  with practically the same approximation of the steady annual cycle. Except for
  a few outliers, the number of model years to obtain a tolerance of $10^{-4}$
  in \eqref{eqn:StoppingCriterion} was nearly identical for the different time
  steps (Fig. \ref{fig:Lhs-RequiredModelYears-RelError}). Both with the
  complexity of the biogeochemical model and with the size of the time step, the
  number of parameter vectors increased for which the spin-up did not reach the
  tolerance of $10^{-4}$ after a maximum of \num{10000} model years (Table
  \ref{table:LHS-Spinup}). If the spin-up did not reach the desired spin-up
  tolerance for a time step, the spin-up using a larger time step in most cases
  did not reach this spin-up tolerance either for the same parameter vector.
  Nevertheless, there were parameter vectors accomplishing the desired spin-up
  tolerance for larger time steps in contrast to smaller time steps. Not any
  parameter vector, for example, existed for which the spin-up reached the
  tolerance of $10^{-4}$ using the NPZD-DOP model. Figure
  \ref{fig:Lhs-RequiredModelYears-RelError} indicates that the tracer
  concentrations, calculated with a spin-up using time steps smaller than
  \SI{8}{\Timestep}, approximated the steady state in the same way for the N,
  N-DOP and MITgcm-PO4-DOP model while the relative error continuously increased
  when time steps larger than \SI{8}{\Timestep} were used. Conversely, the
  deviation from the steady annual cycle of the reference solution grew steadily
  when larger time steps were used for the other three models NP-DOP, NPZ-DOP
  and NPZD-DOP. In particular, the deviations using time step \SI{64}{\Timestep}
  (for all three models) as well as \SI{32}{\Timestep} (for the NPZ-DOP and
  NPZD-DOP model) were noticeable.

\conclusions
\label{sec:Conclusion}

  Using larger time steps to compute steady annual cycles provided practically the
same solution for marine ecosystem models based on the TMM, and shortened the
runtime. We computed steady annual cycles for a hierarchy of biogeochemical
models of increasing complexity \citep[cf.][]{KrKhOs10, DuSoScSt05} using larger
time steps to reduce the computational effort. Apart from that, the solution
computed with a larger time step conformed with the solution calculated with
time step \SI{1}{\Timestep} with a suitable precision. More importantly, we can
mostly apply larger time steps to compute steady annual cycles with a decreased
computational effort.

The error between the reference solution and the solution calculated with a
larger time step increased with larger time steps. This occurred by reason of
the discretization error of the explicit Euler method \citep{StoBul02} and of
the utilization of an approximation of the transport matrix for larger time
steps \citep{Kha07, PiwSla16a}. The aim was to determine the time step as large
as possible so that the discretization error is small enough. For the three
most complex models, for example, the relative error exceeded $10^{-2}$ using
time steps larger as \SI{8}{\Timestep} (cf. Figs. \ref{fig:Norm-2} and
\ref{fig:Lhs-SpinupNorm-RelError}). Hence, the time step must be selected for
every model so that the error is appropriate for the particular application of
the steady annual cycle. However, the accuracy of norm of differences
\eqref{eqn:StoppingCriterion} did not allow to suggest the accuracy of the
solution.

The divergence of the spin-up using larger time steps for the three
biogeochemical models with the highest complexity, we have explained by a too
large step size for the Euler method. As a consequence, the biogeochemical model
ran empty because the model required, for example, more nutrients than were
available in a box of the discretization due to the large time step. As a
result, the concentration became negative. The absolute value of the
concentrations increased during the spin-up so that NaN arose by the computation
of the zooplankton grazing because the numerator as well as the denominator of
the quadratic term became infinite (cf. \eqref{eqn:Zooplankton}). Subsequently,
the occurrence of NaN increased because the result of an operation is NaN if at
least one operand is NaN \citep{Gol91}. The divergence depended, additionally,
on the model complexity and the parameters.

The application of larger time steps lowered the computational costs of the
steady annual cycle calculation for marine ecosystem models. Instead of
\num{2880} time steps per model year using time step \SI{1}{\Timestep}, time
step \SI{2}{\Timestep} merely takes up \num{1440} time steps per model year
and time step \SI{64}{\Timestep} only requires \num{45} time steps per model
year. Therefore, a speed-up factor of up to \num{64} is possible when larger
time steps are used. The TMM easily supports the application of larger time
steps \citep{Kha07} and \citet{PiwSla16a} implemented this application in
Metos3D. Our results provide crucial evidence that the application of larger
time steps practically yielded the same solution. For instance, \citet{PPKOS13}
applied time step \SI{64}{\Timestep} using Metos3D and \citet{KriOsc15} as well
as \citet{KSKSO17} used time step \SI{4}{\Timestep} with the TMM for the
computation of steady annual cycles, respectively.

We have computed the steady annual cycle based on the application of larger time
steps for many different parameter sets in a huge domain for each one of the
models of the model hierarchy. The model assessment \citep[e.g.,][]{KrKhOs10},
the model calibration \citep[e.g.,][]{Kri17} or the sensitivity analysis
\citep[e.g.,][]{KrOsKh12} of global marine biogeochemical models requires
quite a lot of computations of a steady annual cycle. Hence, using larger time
steps lowers the computational cost considerably. Another field of
application for the use of larger time steps is the parameter identification
\citep[for example carried out in][]{PPKOS13}. Accordingly, the runtime can
considerably be shortened using larger time steps because the optimization needs
the computation of a steady annual cycle for several parameter sets. Future work
will be directed to implement an algorithm for the computation of a steady
annual cycle using automatically the largest possible time step, such as a step
size control.

In summary, the main points are the following:
\begin{itemize}
  \item Using larger time steps shortened the runtime of simulations of marine
        ecosystem models.
  \item The computation of a steady annual cycle using larger time steps for
        marine ecosystem models practically yielded the same solution.
  \item A too large time step led to a divergent spin-up calculation, especially
        for complex biogeochemical models.
\end{itemize}


\codedataavailability
{
  The code used to generate the data in this publication is available at 
  \url{https://github.com/slawig/bgc-ann}, \url{https://metos3d.github.io/}
  and \url{https://github.com/jor-/simulation}. We applied version v0.5.0
  of Metos3D - with version v0.2.2 of the Metos3D data package, v0.3.3
  of the Metos3D model package and v0.5.0 of the Metos3D simulation package -
  for all numerical experiments. All used and generated data are available at
  \url{https://doi.org/10.5281/zenodo.5643706} \citep{PfeSla21}.
} 















\bibliographystyle{copernicus}
\bibliography{Timesteps}

\end{document}